\documentclass[11pt,USenglish,pdflatex]{article}
\usepackage{amsmath,amsfonts,amssymb}
\usepackage[font=scriptsize]{subfig}
\usepackage{babel,graphicx}
\usepackage[square,numbers]{natbib}
\usepackage{xcolor}
\usepackage{placeins}
\usepackage{booktabs,colortbl}
\usepackage{doi}
\usepackage{lineno}
\usepackage{todonotes}
\usepackage{enumitem, kantlipsum}
\usepackage[left=2.5cm,right=2.5cm,top=2.5cm,bottom=2.5cm,
headheight=18pt,headsep=18pt,footskip=24pt,asymmetric=true]{geometry}
\usepackage{booktabs}
\usepackage[font=scriptsize]{subfig}
\begin{document}

\title{Multiresolution Coupled Vertical Equilibrium Model for Fast Flexible Simulation of \co Storage}%
   \author{Olav M{\o}yner and Halvor M{\o}ll Nilsen}

\date{\today}

%
\newcommand{\colab}{\co-lab}

\newcommand{\vect}[1]{\boldsymbol{#1}}
\newcommand{\mat}[1]{\boldsymbol{#1}}
\newcommand{\discrete}[1]{\boldsymbol{#1}}
\newcommand{\tens}[1]{\boldsymbol{\mathsf{#1}}}
\newcommand{\transp}[1]{{#1}^{\ensuremath{\mathsf{T}}}}
\newcommand{\K}{\tens{K}}
\newcommand{\mA}{\mat{A}}
\newcommand{\mB}{\mat{B}}
\newcommand{\mD}{\mat{D}}
\newcommand{\mP}{\mat{P}}
\newcommand{\mL}{\mat{L}}
\newcommand{\mR}{\mat{R}}
\newcommand{\vp}{\vect{p}}
\newcommand{\vq}{\vect{q}}
\newcommand{\vc}{\vect{c}}
\newcommand{\vb}{\vect{b}}
\newcommand{\vd}{\vect{d}}
\newcommand{\vx}{\vect{x}}
\newcommand{\vy}{\vect{y}}

\newcommand{\co}{CO$_2$ }

\newcommand{\vevar}[1]{\MakeUppercase{#1}}
\newcommand{\ddens}{\discrete{\rho}}
\newcommand{\dmob}{\discrete{\lambda}}
\newcommand{\dsat}{\discrete{s}}
\newcommand{\dq}{\discrete{q}}
\newcommand{\dpres}{\discrete{p}}
\newcommand{\dz}{\discrete{z}}
\newcommand{\dv}{\discrete{v}}

\newcommand{\upw}{\mbox{\texttt{upw}}}
\newcommand{\favg}{\mbox{\texttt{favg}}}
\newcommand{\grad}{\mbox{\texttt{grad}}}
\newcommand{\sgn}{\mbox{\texttt{sign}}}

\newcommand{\gas}{{g}}
\newcommand{\water}{{w}}
\newcommand{\reseq}{\mathcal{R}}

\newcommand{\gacc}{\Theta}
\newcommand{\gsat}{\Pi}
\newcommand{\gq}{\Psi}
\newcommand{\Fine}{\mathrm{Fine}}
\newtheorem{example}{Example}
\newcommand{\rev}[1]{\textcolor{red}{#1}}

\renewcommand{\topfraction}{0.95}	
\renewcommand{\bottomfraction}{0.8}	
\setcounter{topnumber}{2}
\setcounter{bottomnumber}{2}
\setcounter{totalnumber}{2}     
\setcounter{dbltopnumber}{2}    
\renewcommand{\dbltopfraction}{0.9}	
\renewcommand{\textfraction}{0.02}	
\renewcommand{\floatpagefraction}{0.9}	
\renewcommand{\dblfloatpagefraction}{0.7}	


\maketitle

\begin{abstract}
	\co capture and storage is an important technology for mitigating climate change. Design of efficient strategies for safe, long-term storage requires the capability to efficiently simulate processes taking place on very different temporal and spatial scales.  The physical laws describing \co storage are the same as for hydrocarbon recovery, but the characteristic spatial and temporal scales are quite different. Petroleum reservoirs seldom extend more than tens of kilometers and have operational horizons spanning decades. Injected \co needs to be safely contained for hundreds or thousands of years, during which it can migrate hundreds or thousands of kilometers. Because of the vast scales involved, conventional 3D reservoir simulation quickly becomes computationally unfeasible. Large density difference between injected \co and resident brine means that vertical segregation will take place relatively quickly, and depth-integrated models assuming vertical equilibrium (VE) often represents a better strategy to simulate long-term migration of \co in large-scale aquifer systems. VE models have primarily been formulated for relatively simple rock formations and have not been coupled to 3D simulation in a uniform way. In particular, known VE simulations have not been applied to models of realistic geology in which many flow compartments may exist in-between impermeable layers. In this paper, we generalize the concept of VE models, formulated in terms of well-proven reservoir simulation technology, to complex aquifer systems with multiple layers and regions. We also introduce novel formulations for multi-layered  VE models by use of both direct spill and diffuse leakage between individual layers. This new layered 3D model is then coupled to a state-of-the-art, 3D black-oil type model. The formulation of the full model is simple and exploits the fact that both models can be written in terms of generalized multiphase flow equations with particular choices of the relative permeabilities and capillary pressure functions. The resulting simulation framework is very versatile and can be used to simulate \co storage for (almost) any combination of 3D and VE-descriptions, thereby enabling the governing equations to be tailored to the local structure. We demonstrate the simplicity of the model formulation by extending the standard flow-solvers from the open-source Matlab Reservoir Simulation Toolbox (MRST), allowing immediate access to upscaling tools, complex well modeling, and visualization features. We demonstrate this capability on both conceptual and industry-grade models from a proposed storage formation in the North Sea. While the examples are taken specifically from \co storage applications, the framework itself is general and can be applied to many problems in which parts of the domain is dominated by gravity segregation. Such applications include gas storage and hydrocarbon recovery from gas reservoirs with local layering structure.
\end{abstract}

\section{Introduction}
Most climate mitigation scenarios require large-scale \co storage as a part of carbon capture and sequestration (CCS) to mitigate global temperature increases beyond the target of two degrees Celsius. To be successful, large volumes of \co must be safely stored for thousands of years. Versatile simulation tools play a vital role to optimize utilization of storage resources and ensure safe operations, but also play an integrated part in post-injection monitoring. In most cases, this means simulating a wide variety of probable scenarios, which means that simulators must be accurate, efficient, and robust. Standard simulators developed primarily for hydrocarbon recovery do unfortunately not always meet these three criteria because of the vast differences in physical scales involved in the study of long-term migration of \co in large-scale aquifer systems. The goal of this work is to present a new multiresolution framework in which the granularity of the flow model adapts to the nature of the flow physics to accelerate overall simulation time.  More specifically, we propose to combine traditional flow models in regions where the flow has a complex 3D structure, with effective vertically-integrated models  \cite{NordbottenCelia} in regions where the flow is primarily driven by buoyancy and consists of one or more layers of vertical equilibria.  By use of a fully implicit formulation for the discrete coupled system of equations, we ensure efficient simulation without sacrificing the robustness of state-of-the-art, well-proven simulation methods from the oil and gas industry.

During the injection process, we believe it is best to use standard methods for 3D reservoir simulation to describe the flow in the near-well region. At high-flow rates, the flow is often dominated by viscous forces and will generally have a 3D structure with comparable characteristic time scales in the lateral/vertical directions, which makes the assumption of vertical-segregation less valid. Secondly, we want to utilize extensive work on advanced well-facility modeling done for hydrocarbon recovery. In the near-well regions, our framework therefore utilizes standard 3D black-oil model with two-point flux approximations and mobility-upwind discretization (TPFA--MUP) \cite{brenier1991upstream}, which is what is used in most commercial simulators \cite{eclipse}.

 However, for segregated flow -- which is expected in regions with long lateral distances and time horizons -- depth-integrated models assuming vertical equilibrium have been shown to more efficient \cite{Bandilla:14,CBNB15:wrr} and in many cases more accurate \cite{Bandilla:14,CBNB15:wrr}. Gains in accuracy can be attributed to a more accurate description of the \co--water interface in the strongly segregated case \cite{LiNi10:ecmor,Nilsen:ghgt10} In addition, the VE description is less sensitive to the choice of discretization. Recent extensions of the VE framework have focused on incorporating the most relevant physical effects for \co storage into VE formulations. Early VE models used for \co storage problems assumed a sharp interface between \co and brine and were posed in simple domains \cite{Nordbotten:06,Hesse:08,Gasda:09}. Later, these models have been extended with compressibility \cite{Andersen:13}, convective dissolution \cite{Gasda:11,Mykkeltvedt:12}, capillary fringe \cite{ND11:wrr}, retardation effects caused by small-scale variations in caprock topography \cite{Nilsen12:ijggc,Gasda12:wrr,Gasda13:awr}, various hysteretic effects \cite{Elsa:13,Doster:12,Doster:13}, and heat transfer \cite{Gasda13:ep}. Whereas most studies focus on depth-integration across a single rock layer, a few attempts have been made at simulation relatively simple examples of systems with multiple geological layers \cite{Bandilla12:cvs,Bandilla12:ijggc}. This family of depth-integrated models has also been extended to cases in which full segregation is not achieved, in particular with geological layering in the vertical direction \cite{Guo14:wrr}. For such cases, the computational method is similar to gravity-splitting approaches used e.g., for streamline simulation \cite{Batycky1997,Froysa2000}, but with a 2D pressure equation.

Most of the extensions mentioned above were first demonstrated with little emphasis on exploiting know-how from robust simulation technology developed in the oil and gas industry. In a series of papers  \cite{MRST:co2lab,co2lab:part1,co2lab:part4,Lie15:compgeo,Andersen:tccs8}, modern VE methods for \co sequestration were reinterpreted in the classical formulations developed during the early period of reservoir simulation \cite{Martin:58,Coats:67,Martin:68,Coats:71}, when time-step robustness and mass-conservation aspects using a limited computational budget was the primary focus. Similar formulations were also introduced in the commercial  Eclipse simulator to model thin layers of oil in the Troll field \cite{Henriquez1992}. The resulting formulation is an extension of the traditional pseudo-relative permeability and capillary pressure \cite{Jacks:73,KyteBerry:75,Stone:91,BarkerThibeau:97}, which unifies the 3D and VE discretizations in a generalized model of multiphase flow with more complex functional dependencies than in traditional models. It also provides a uniform framework which is suitable fore TPFA--MUP type discretization.

Our main contribution herein is to make the VE framework fully integrated in a general geological model. In particular, we show coupled, multilayered VE simulations on real formations for the first time, couple 3D and VE in a uniform and stable simulation framework, and introduce simple, extendable diffuse-leakage coupling between VE layers, which is fully consistent with the underlying 3D description of the simulation model.

\co storage is not economically feasible without a global market environment in which the cost of greenhouse gas emissions is incorporated. At the time of writing, this has not yet been achieved, but the simulation infrastructure should be available and proven well before commercial large-scale CCS becomes common. Given the time-constraint implicit in the climate challenge, we believe that the only way robust and novel simulation software for the \co storage problem can be developed and proven beyond doubt, is through collaboration and use of community software. The new framework is therefore developed by use of the Matlab Reservoir Simulation Toolbox (MRST) \cite{MRST:2017a,MRST12:comg,Lie:mrst-book,mrst-ad:rss15}, which is an open-source community software that can be freely downloaded and used under the GNU General Public License v3.0. The software offers reliable simulations of models with full industry-standard complexity, and  enables interactive experimentation with various model assumptions like boundary conditions, different fluid models and parameters, injection points and rates, geological parameters, etc. The software also offers a wide variety of computational methods the user can combine to quickly create models and tools of increasing fidelity and computational complexity for modeling \co storage, e.g., as discussed by \cite{Lie15:compgeo,Andersen:tccs8}. These tools have previously been applied to study ongoing injections at Sleipner, as well as many synthetic scenarios posed on geological models of saline aquifers from the Norwegian Continental Shelf \cite{CO2atlas:NCS}.

Our new framework will be constructed based on several modules from MRST. By use of functionality for upscaling and grid coarsening, we produce user-friendly workflow that takes a fine-grid 3D simulation model as input and automatically converts it to a coupled 3D-VE model, with different regions automatically determined by the simulator. The only additional input is a specification of which domains should be treated using 3D or VE discretizations. The hybrid 3D--VE grid is formulated as a simple partition of the underlying grid, and hence it is straightforward to also develop strategies in which the vertically averaged regions adapt dynamically to the migrating \co plume, as recently proposed by \citep{Helmig:17siamgs}; dynamic coarsening methods have previously been demonstrated in MRST, e.g., by \cite{HLN12:comg}. To build a hybrid 3D--VE flow simulator, we utilize a combination of standard black-oil equations, which have been implemented using automatic differentiation \cite{mrst-ad:rss15,Lie:mrst-book}, with functionality from MRST-co2lab for computing dynamic pseudo functions based on depth-integration and an assumption of vertical equilibrium \cite{co2lab:part2,co2lab:part3}. We demonstrate our new framework on geological models from the Utsira formation in the Norwegian North Sea, and verify large efficiency gains compared with fine-scale simulation. In fact, the gains are much larger than what can be explained by the mere reduction in the number of degrees-of-freedom, which is consistent with observations for simpler cases \cite{Bandilla:14,CBNB15:wrr}.

The implementation described in this paper is not yet publicly available, but will be included in an upcoming release of MRST. Using this general framework opens up for many possible extensions of the work presented herein. Use of automatic differentiation makes it straightforward to compute gradients and parameter sensitivities, e.g., by use of an adjoint formulation. This enables users to easily perform optimization and sensitivity studies with respect to parameters of interest, which is essential to investigate optimal injection strategies and develop closed-loop strategies for \co storage management. Such strategies consist of regular assimilation of measured data, and could require a coupling of flow simulations with, for example, time-dependent seismic, gravity, or other monitoring data. We have previously demonstrated the usefulness of sensitivities for pure VE model used as part of rigorous mathematical optimization of large-scale injection strategies \cite{Andersen14:ecmor,Lie14:ecmor,Lie15:compgeo,co2lab:part4} or to match models with interpreted time time-lapse seismics \cite{Nilsen2017}. Finally, we mention that our new framework has been implemented using object-oriented techniques which should simplify subsequent extensions to include additional physics such as thermal, geochemical, and geomechanical effects, which are already available in the software.

\section{Governing equations}
To describe our method, we consider two immiscible phases that each contain a single component: the aqueous phase contains resident brine, whereas the vapor/supercritical phase contains the \co component.
use two-component, two-phase flow in a porous medium. The governing equations for immiscible, two-phase flow can be written as mass-conservation equations for each component or phase,
\begin{align}
\label{eq:govern}
\frac{\partial}{\partial t}(\phi s_\water \rho_\water)& + \nabla \cdot (\rho_\water \vec{v}_\water) = q_\water ,\\ \frac{\partial}{\partial t}(\phi s_\gas \rho_\gas)& + \nabla \cdot (\rho_\gas \vec{v}_\gas) = q_\gas ,
\end{align}
where $\rho_\alpha(p)$ is the density of a phase, $\phi(p)$ the fraction of the medium available to flow, $s_\alpha$ the fraction of the open pores occupied by phase $\alpha$, and $q_\alpha$ source terms. Phase velocity is determined using the standard multiphase extension of Darcy's law,
\begin{equation}
\label{eq:darcy}
\vec{v}_\alpha = - \lambda_\alpha \K \nabla (p_\alpha - \rho_\alpha g \Delta z), \quad \lambda_\alpha = \frac{k_{r\alpha}}{\mu_\alpha},
\end{equation}
where the relative permeability $k_{r\alpha}(s_\alpha)$ models reduction in flow rate due to the presence of the other phase as a function of saturation, $\mu_\alpha$ is the fluid viscosity, and $g$ is the gravity constant acting in the vertical direction for increasing depth. The phase mobility $\lambda_\alpha$ is the ratio between relative permeability and viscosity. This equation, in addition to modified functions relating the phase pressures, will be generalized to account for the new models and couplings.

To close the system, we assume that the two phases fill up the available pore-volume,
\begin{equation}
\label{eq:satclose}
s_\water + s_\gas = 1,
\end{equation}
and that the difference between the phase pressures is given by the capillary pressure as a function of saturation,
\begin{equation}
\label{eq:capillary}
p_\gas = p_\water + p_{wg} (s_\gas).
\end{equation}
We emphasize that the two-component immiscible flow was chosen to simplify the description of our new framework. The underlying implementation seems to work equally well for more complex black-oil models as well as a general compositional model with miscibility.

\section{Discretization}
In this section, we describe the discretization of the model equations. We begin by briefly discussing the fine-scale and vertical equilibrium on discrete form that are used, before considering the transition between different discretization regions, which is the main objective in this paper. Although the fine-scale and VE equations should be fairly well-known to the reader, the specific notation used herein is useful for the treatment of the coupling terms later on.
\subsection{Fine-scale}
For a fine-scale discretization of the mass-flux (\ref{eq:govern} multiplied by density), we employ a standard two-point flux approximation together with first-order mobility upwinding \cite{brenier1991upstream}. The temporal derivative is discretized using backward Euler. For a given cell index $i$ with a set of neighboring cell indices $N(i)$, we then obtain,
\begin{equation}
\label{eq:discrete_fine}
\frac{1}{\Delta t}\left[ (\Phi_i \dsat_\alpha \ddens_\alpha)^{n+1}_i + (\Phi_i \dsat_\alpha \ddens_\alpha)^{n}_i \right] - \sum_{j \in N(i)}f_{ij}(\dpres_\alpha^{n+1}, \ddens_\alpha^{n+1}, \dmob_\alpha^{n+1}, \dz) = (\dq_\alpha)_i^{n+1}.
\end{equation}
In the flux, we have used $\dz$ as the depths of the cell-centers. If we use a standard two-point approximation for the fluxes \eqref{eq:darcy}, we can obtain discrete phase flux given for an oriented interface $ij$ as,
\begin{equation}
\label{eq:discflux}
f_{ij}(\dpres_\alpha, \ddens_\alpha, \dmob_\alpha, \dz) = -\, \upw(\ddens_\alpha \dmob_\alpha, \dpres_\alpha)_{ij} T_{ij}\left[\grad(\dpres)_{ij} -  g\, \favg(\ddens_\alpha ) \grad (\dz)_{ij}  \right].
\end{equation}
We remark that the mobility $\lambda_i^{ij}$ can be considered as one value for each oriented face. This generalization is equivalent to considering tensor relative permeability in the discrete setting of a two point flux mobility upwind discretization. For brevity, we neglect rock compressibility in this discussion and assume that the source terms are known functions of pressure, saturation and time. By eliminating one saturation and one pressure using the closure relations \eqref{eq:satclose} and \eqref{eq:capillary} this forms a well-posed system of equations with one phase pressure and one phase saturation as primary variables.

We have introduced discrete operators $\grad$, $\favg$ and $\upw$ to map values from cells to faces. For the transported quantities in the flux expression, we use the first-order upwind operator, which is defined for a given cell-wise vector $\dv$ and the discrete phase pressure $\dpres$ as,
\begin{equation}
\upw(\dv, \dpres)_{ij} =
\begin{cases}
	\dv_i \mbox{ if } \dpres_i > \dpres_j \\
	\dv_j \mbox{ otherwise.}
\end{cases}
\end{equation}
The discrete gradient $\grad$ and the face average $\favg$ are, respectively, the difference in values over the interface and the averaged value of the two cell values,
\begin{equation}
\grad(\dv)_{ij} = \dv_j - \dv_i, \qquad \favg(\dv)_{ij} = \frac{\dv_i + \dv_j}{2}.
\end{equation}

\subsection{Vertical equilibrium}
The vertical equilibrium (VE) approach is derived assuming that buoyancy is a significantly stronger force than the viscous forces at the chosen grid resolution and aspect ratio. While this assumption is not valid for all flow scenarios, it is well suited for simulation of \co storage due to the significant density difference between brine and \co as well as the long temporal and spatial scales that result in strong gravity segregation and flow regimes that are often driven by gravitational forces.
\begin{figure}[ht]
	\centering
	\includegraphics[width = 1\textwidth]{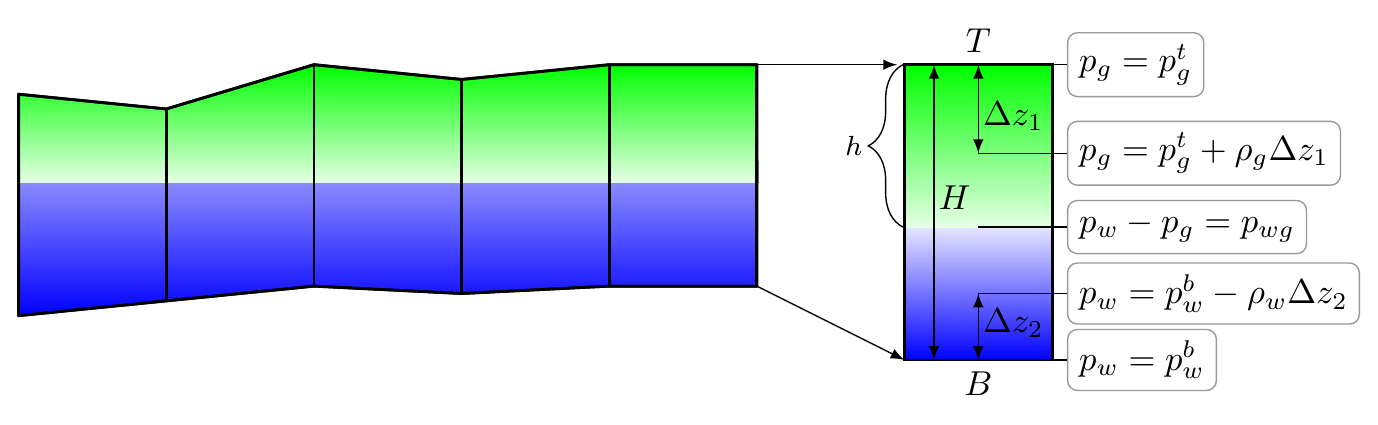}
	\caption{Water and gas pressures inside a vertically segregated column with height $H$ and constant densities.}
	\label{fig:show_ve}
\end{figure}
In the case where the fine-scale capillary forces is negligible, the \co will be above the water and the \co-water contact will be a sharp interface. We define the height $H$ for each cell and assume complete hydrostatic segregation within the cell such that the \co is located as a column with height $h$ from the top of the cell as shown in Figure~\ref{fig:show_ve}. If we assume the aqueous pressure at the bottom of the cell, $p_w^b$ to be known as well as the vapor pressure at the top of the cell $p_g^t$, we can find the pressures at any point inside the aqueous and vapor columns using the constant density of each phase within the cells. These pressures must be equal to the entry pressure at the interface between the two phases. We define the vapor and aqueous pressures in the VE-model at the top and bottom of the column, respectively. The pressure difference due to the buoyancy force exerted by the column of \co takes a form analogous to the fine-scale capillary force, resulting in the final pressure difference between the two phases in a VE zone as,
\begin{equation}
\vevar{p}_\gas = p^t_\gas, \, \vevar{p}_\water = p^b_\water, \qquad \vevar{p}_\gas = \vevar{p}_\water + p_{wg}(\vevar{S}_g) + h g (\rho_\water - \rho_\gas) - Hg\rho_\water.
\end{equation}
This expression is found by inserting the definitions of VE pressures and solving for equal phase pressure at the fluid interface.

We have opted to use capital case to refer quantities under the VE assumption to differentiate them from the fine-scale. In the above case there is now flowing water at the top and $P_w$ is equal to the hydrostatic head of water at the bottom with respect the water pressure at the \co-water interface. Assuming a level, sharp interface between two phases, the height of the \co column can be found from the saturation as $h = H \vevar{s}_g$ and the relative permeability will be equal to the saturation. In more advanced models, for example models where capillary pressure modifies the interface, the relations are more complicated, see e.g. \cite{co2lab:part3}. We can now write the discrete VE system on the same form as \eqref{eq:discrete_fine}, albeit with a slightly different interpretation of the variables,
\begin{eqnarray}
\label{eq:discrete_ve}
\frac{1}{\Delta t}\left[ (\Phi_i \vevar{\dsat}_\water \ddens_\water)^{n+1}_i + (\Phi_i \vevar{\dsat}_\water \ddens_\water)^{n}_i \right] - & \sum \limits_{j \in N(i)}f_{ij}(\vevar{\dpres}_\water^{n+1}, \vevar{\ddens}_\water^{n+1}, \vevar{\dmob}_\water^{n+1}, \dz^b) = (\vevar{\dq}_\water)_i^{n+1},\\
\frac{1}{\Delta t}\left[ (\Phi_i \vevar{\dsat}_\gas \ddens_\gas)^{n+1}_i + (\Phi_i \vevar{\dsat}_\gas \ddens_\gas)^{n}_i \right] - & \sum \limits_{j \in N(i)}f_{ij}(\vevar{\dpres}_\gas^{n+1}, \vevar{\ddens}_\gas^{n+1}, \vevar{\dmob}_\gas^{n+1}, \dz^t) = (\vevar{\dq}_\gas)_i^{n+1}.
\end{eqnarray}
In the above expression, $\dz^t$ and $\dz^b$ correspond to the top and bottom depths of the coarse cells. This is slightly different from the definitions in \cite{co2lab:part2,co2lab:part3} where only a single region of VE was considered. This definition is simpler to use in the general setting because the flow of the different phases are defined at the point where they physically appear: The top, $z^t$, for \co and the bottom, $z^b$, for water. The only complication is that the pressure at the point has to be reconstructed from the variables defined as primary. In this work, we use the cell-center pressures for both primary variables and property evaluations, requiring a small hydrostatic correction before fluxes are computed.
\subsection{Discrete fluxes at the interfaces}
\begin{figure}[ht]
	\centering
	\subfloat[Transition from VE to fine]{
		\includegraphics[width=.3\textwidth]{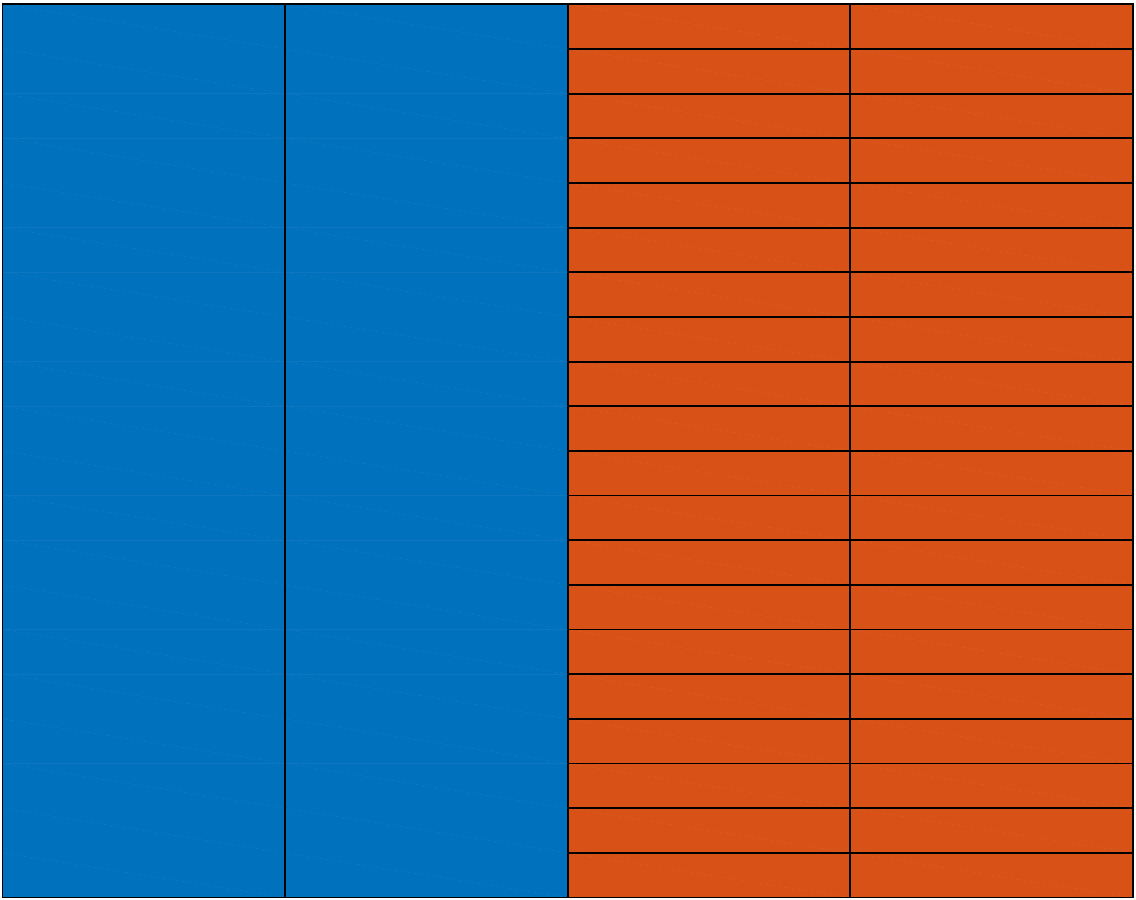}} \hspace{1cm}
	\subfloat[Transition from VE to VE]{
		\includegraphics[width=.3\textwidth]{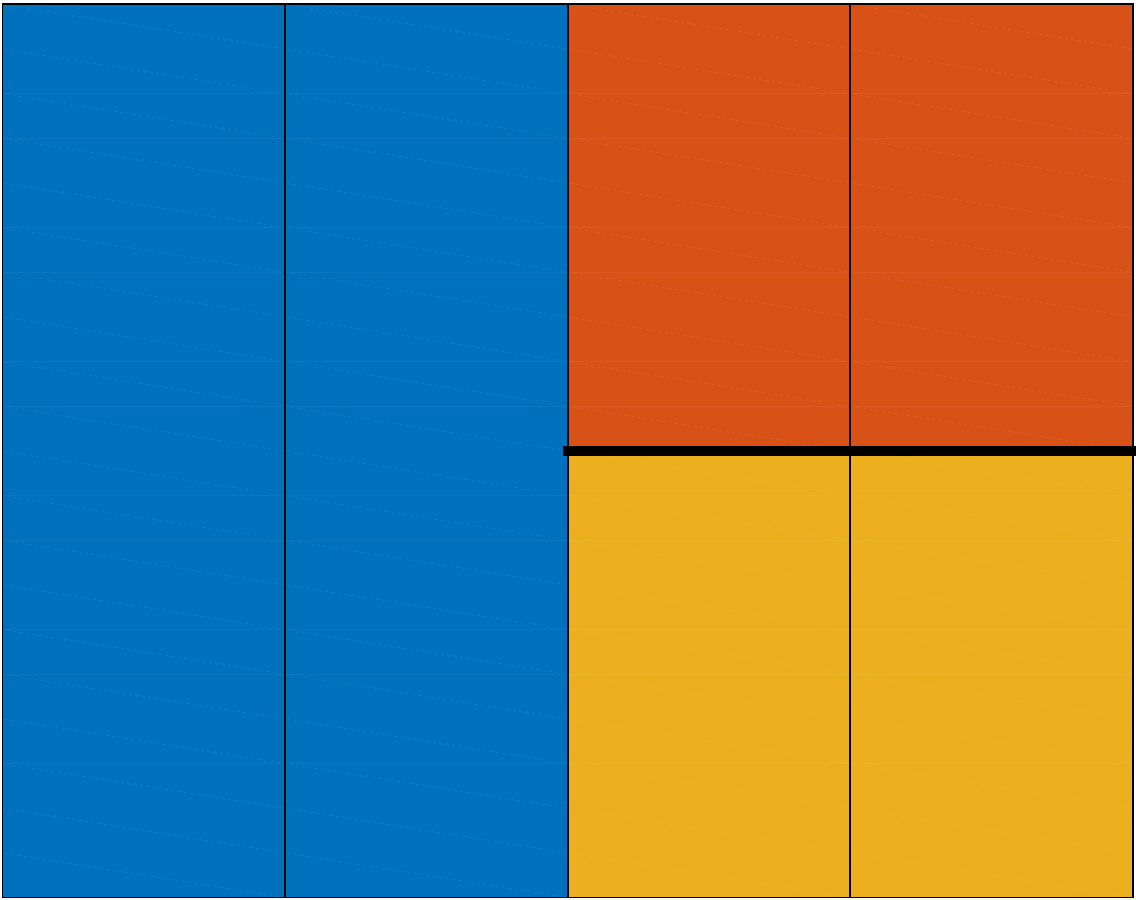}}
	\caption{The two transition cases: From a zone with a VE discretization to a zone with a fine-scale discretization (left) and the transition from a single to two different VE zones, separated by a impermeable layer.}
	\label{fig:transitions}
\end{figure}

We have in the previous sections introduced both a fine-scale and an upscaled system in a discrete setting. The VE approach is in general valid for cells where the gravity segregation due to buoyancy happens on a much shorter time-scale than the transport induced by viscous forces. The most restrictive of the VE assumptions are that of full gravity segregation, see \cite{Guo14:wrr}
for a discussion of a more advanced model. VE is well suited for the simulation of \co migration, where the time-scales are long and there exists a significant density difference. However, the formulation assumes a smooth top surface without significant vertical flow. This situation may occur, e.g. in the presence of partially eroded impermeable shale layers between two regions of the reservoir.

In order to create a VE model fully suitable for general geomodels, we wish to both be able to couple different VE regions characterized by a discontinuous top surface. In addition, the assumption of vertical equilibrium will not be valid in high flow regions, for example near wells. In these parts of the domain, it may be required to use a fine grid resolution to resolve the flow pattern. To overcome these challenges, we need expressions for $f_{ij}$ at interfaces where $i$ and $j$ belong to different \emph{discretization regions}, either representing the transition between two VE zones or from one VE zone to a fine-scale grid cell. Both cases are represented in Figure~\ref{fig:transitions}
\subsubsection{Virtual cells with local saturations and pressure}
Since the accumulation terms representing total mass in a cell and the source terms are independently defined per cell, transitioning between two discretizations comes down to defining the numerical flux function \eqref{eq:discflux} at the interface. In order to provide a relatively generic implementation, we will not modify the flux function directly, but rather choose the \emph{values} of $p,\rho,\lambda,z$ carefully.
\begin{figure}[ht]
	\centering
	\includegraphics[width = .4\textwidth]{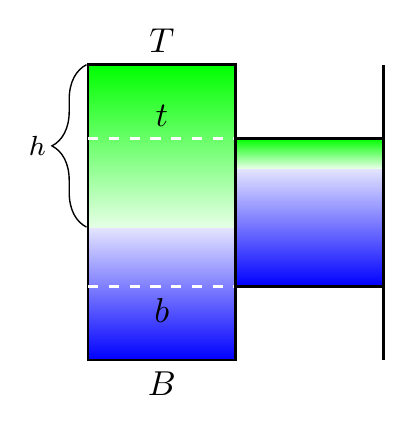}
	\caption{Transition zone between two discretization regions. The left column has a higher edge than the right column, resulting in a non-matching interface. To overcome this, a virtual cell with a matching interface is constructed. The virtual cell next to the interface is shown in white.}
	\label{fig:virtual_cell}
\end{figure}
Figure~\ref{fig:virtual_cell} shows the fine details near a single transition interface, between two VE cells corresponding to different layers. In this case, we will focus on the values in the left cell specifically as the treatment of the right cell is analogous. In order to keep the discussion as simple as possible, we assume that pillars are completely vertical with a flat top and bottom surface. For a given pillar, we have $z^t$ and $z^b$ as the z-depth of the top and bottom of the column, respectively. For a given interface between the column and a neighboring column, we define $z^t_{i}$ and $z^b_{i}$ for the column as the top and bottom of the neighboring column\footnote{In our implementation, we exploit the fact that the grid is a coarse grid where the cells consist of agglomerations of fine-cells. We look up the fine-scale cells at the interface, from which $t$, $h$ can be retrieved directly. The pre-processor does not need to take geometry into account, it simply uses the fine-grid depths.}.

The region inside our column delineated vertically by $t$ and $h$ defines a \emph{virtual cell}. If we can determine the saturations and pressure in this cell for both the fine-scale and VE discretizations, we can use the flux function directly to obtain a consistent flux. Since the parent cell of the virtual cell is assumed to be in vertical equilibrium with a sharp interface, we can write the virtual gas saturation $(s_g)_v$ as a function of the plume height. This value is the same if the cell is interpreted as a fine-scale or a VE-cell, given using positive direction along increasing depth,
\begin{equation}
(s_g)_v = (\vevar{s}_g)_v =
\begin{cases}
0 & \mbox{ if } T + h \leq t \\
(h - t + T)/(b-t) & \mbox{ if } t < T + h < b \\
1 & \mbox{ if } T + h \geq b.
\end{cases}
\end{equation}
We also require the fine-scale water pressure at a point $x$, $(p^x_w)_v$. This value inside the virtual cell is found by integrating the hydrostatic pressure distribution from the known pressure $\vevar{p}_w$ at the bottom of the column. Let the shallowest depth of the gas plume and the point of interest $x$ inside the virtual cell be $d = \min(T+h, x)$ to obtain
\begin{align}
\label{eq:virtual_pressure}
(p^x_w)_v = & \vevar{p}_w - \int \limits_B^x (\rho_w s_w + \rho_g s_g) g \, d\xi\\ =& \vevar{p}_w - g \left[\int \limits_B^d \rho_w  \, d\xi + \int \limits_d^{x} \rho_g  \, d\xi \right]\\= & \vevar{P}_w - g(B-d)\rho_w - g(d-x)\rho_g,
\end{align}
where we have used that the phase densities are constant within each fluid column. If we set $x = b$, we obtain $(\vevar{p}_w)_v$. For the fine-scale pressure defined at the center of the cell, $p_w$ it is sufficient to set $x = (b + t)/2$. We can now obtain both the pressure and the saturation in the virtual cell at the interface in both the VE and fine-scale variable sets. 

\subsubsection{Vertical equilibrium transition zone}
In the transition between two vertical-equilibrium zones, we reconstruct local heights, saturations and depths in the virtual cell corresponding to the intersection between the two columns and treat the connection as VE-to-VE interface. We note that this choices gives a different form than the expression obtained for the VE-to-VE for two cells of the same region, where the transition is defined in terms of the saturations of the whole column. This is because the fine-scale situation is different. In the VE-to-VE interaction using the virtual cell we have assumed that the height difference between the columns are only due to a discontinuity, while in the VE-to-VE case of same layer the top surface is assumed to be smoothly varying. These two interpretations can lead to different large scale behavior and can be related to the work in \cite{Gasda12:wrr} on upscaling of top-surface topography. The two expressions represent the two limit cases of a smooth family of upscaled relative permeabilities valid in different regimes. For example, the smooth version is the correct for the discontinuous top surface in the limit where the viscous forces dominate buoyancy forces and co-current flow dominates the counter-current segregation effects. For large scale simulations, the virtual cell treatment is appropriate where the model is expected to have a discontinuity in the surface, e.g. over connecting fault lines. For VE-to-VE where more than one VE cells is involved on one side the virtual cell concept is the only physically relevant approach, as any other treatment would allow for larger-than-unity relative permeability on the coarse-scale.

\subsubsection{Fine-scale to VE transition}
Cells connected to a transition zone which represent fine-scale cells have pressure defined at the centroids rather than the top of the cell and we require an additional pressure correction. In such cells we change the starting point of the integration in equation \eqref{eq:virtual_pressure} to the cell centroid instead of the top of the column. For purposes of the transition from fine to VE zones, the virtual cells belonging to vertically-equilibrated columns are treated as fine-scale cells, with reconstructed pressure and saturations. The relative permeability is evaluated using the reconstructed fine-scale saturation.

\subsubsection{Vertical flow between VE regions}
Certain aquifer models contain nearly impermeable layers that compartmentalize the flow. These layers are of particular interested for CO$_2$ storage, as the assumption of gravity segregation within each column for long time-scales is challenged by almost impermeable layers. For connections between two VE-regions in the vertical direction, we reconstruct fine-scale virtual cells over the layer and treat the flux as between two fine-scale cells. Specifically, this is required in the Utsira example, where the model uses translatability multipliers to account for diffusive leakage through thin clay layers. This connection is treated separately in our implementation, allowing for the rapid incorporation of more advanced diffuse leakage models. 

\subsection{Coupled equations}
In the preceding text, we used the $v$ subscript to refer to a virtual cell since the discussion was limited to a single interface. For the general discretization, we employ the notation $\dv_i^{ij}$ to refer to the virtual value of the discrete variable $\dv$ in a cell $i$ when considered at the interface $ij$. We assume that this definition is unique per interface, $\dv_i^{ij} = \dv_i^{ji}$. The virtual cell always belongs to a single coarse block-interface pair. We also need the notion of a column, i.e. a set of cells that are vertically layered on top of each other, is found using the indicator $C(i)$ which provides the column-index of cell $i$.

We define the discrete indicator $V(i)$ which gives the ve-region of cell $i$. If a cell is not governed by the VE discretization, $V(i)=$ Fine. We can write the discretization of the coupled system as,
\begin{equation}
\label{eq:discrete_global}
\frac{1}{\Delta t}\left[ \gacc^{n+1}_i + \gacc^{n}_i \right] - \sum_{j \in N(i)}f_{ij}(\gsat_\alpha^{n+1}) = \gq_i^{n+1},
\end{equation}
where we change accumulation terms based on the discretization for the individual cells,
\begin{equation}
\gacc_i =
\begin{cases}
\Phi_i \dsat_i \ddens_i \mbox{ if } V(i) = \Fine,\\
\Phi_i \vevar{\dsat}_i \vevar{\ddens}_i \mbox{ otherwise,}\\
\end{cases} \quad
\gq_i =
\begin{cases}
\dq_i \mbox{ if } V(i) = \Fine,\\
\vevar{\dq}_i \mbox{ otherwise.}\\
\end{cases}
\end{equation}

There are four categories of flux variables to consider. We have already discussed the case for $V(i) = V(j) = \Fine$, i.e. transition between two fine-scale cells. In the same vein, we have discussed the case where two VE-columns of the same category cells exchange mass, $V(i) = V(j) \neq \Fine$. The two remaining cases are the transitions between mixed categories $V(i)\neq V(j)$. Depending on the context, this transition is either treated as a fine-scale or a VE transition. In total, we have five different cases to consider:
\begin{enumerate}[nosep]
	\item Coupling between fine cells.
	\item Coupling between VE cells in the same VE layer: Figure~\ref{fig:show_ve}
	\item Coupling between two VE-columns of different category (VE-scale): Figure \ref{fig:transitions}, right subplot, blue to orange cells or Figure \ref{fig:virtual_cell}.
	\item Coupling in the vertical direction due to diffuse leakage (fine-scale): Figure \ref{fig:transitions}, right subplot, orange to yellow cells
	\item Coupling between fine-scale and VE-columns where $C(i)=C(j)$ but $V(i)\neq V(j)$ (fine-scale)
	\ref{fig:transitions}, left subplot, blue to orange cells.
\end{enumerate}

We reconstruct the appropriate virtual values and apply the same flux expression for all cases. The concise definition of the variables that must be reconstructed for all possible interfaces can thus be written out,
\begin{equation*}
\label{eq:coupled_flux}
\gsat_{ij} =
\begin{cases}
(\dpres, \ddens, \dmob, \dz) &\mbox{ if } V(i) = V(j) = \Fine \mbox{: Case 1,}\\
(\vevar{\dpres, \ddens, \dmob, \dz}) &\mbox{ if } (V(i) = V(j)) \land (C(i) \neq C(j)) \land (V(i) \neq \Fine) \land (V(j) \neq \Fine) \mbox{: Case 2,}\\
(\vevar{\dpres, \ddens, \dmob, \dz})^{ij} &\mbox{ if } (V(i) \neq V(j)) \land (C(i) \neq C(j))\land (V(i) \neq \Fine \land V(j) \neq \Fine)\mbox{: Case 3,}\\
({\dpres, \ddens, \dmob, \dz})^{ij} & \mbox{otherwise: Cases 4 \& 5.}

\end{cases}
\end{equation*}
Note that the coarse depth $\vevar{\dz}$ is assumed to take the top or bottom depending on the phase being evaluated.

\subsection{Automated selection of different regions}
\begin{figure}[ht]
	\centering
	\subfloat[Fine-scale grid with impermeable layers before coarsening.]{
		\includegraphics[width=.3\textwidth]{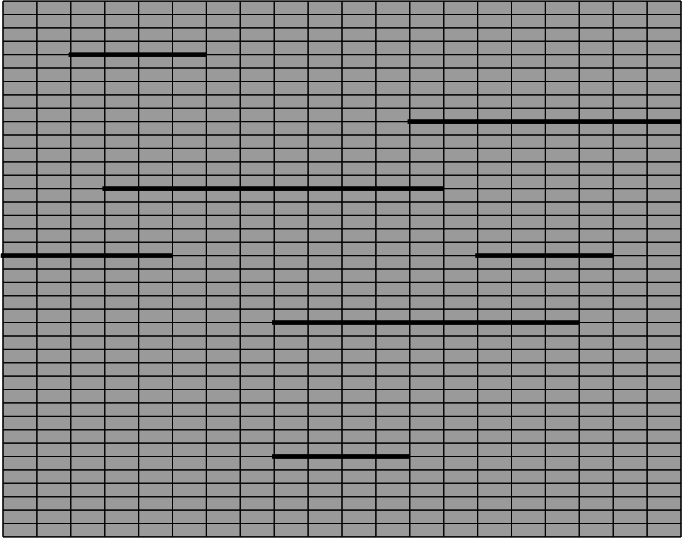}} \hspace{.5cm}
	\subfloat[Columns split into connected components in the vertical direction.]{
		\includegraphics[width=.3\textwidth]{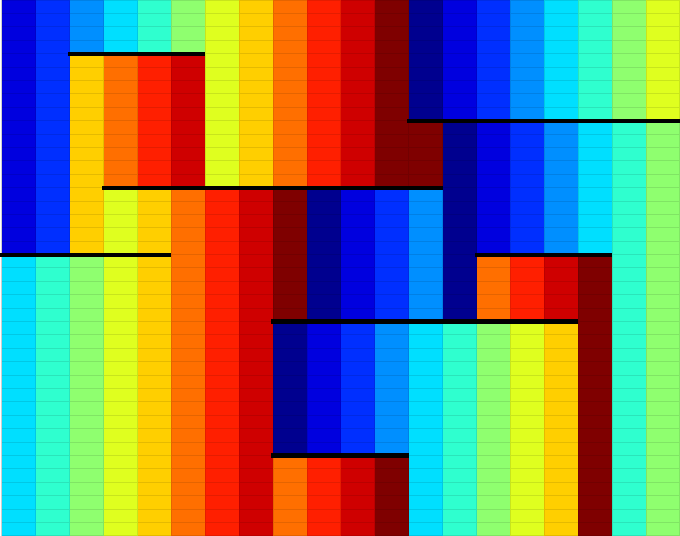}} \hspace{.5cm}
	\subfloat[Blocks with single column connection merged together. These blocks share top and bottom surface.]{
		\includegraphics[width=.3\textwidth]{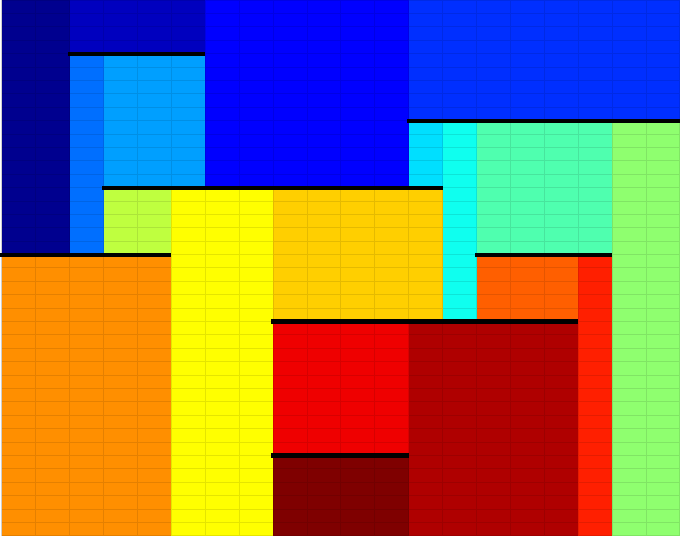}} \hspace{.5cm}
	\caption{The grid with impermeable layers (left) is processed into connected sub-columns (middle) before all sub-columns with the same connectivity are assigned to the same category (right). Each color corresponds in this final figure to a distinct region with uniform discretization.}
	\label{fig:coarsening}
\end{figure}

Applications of the proposed multi-region VE model to realistic reservoir models necessarily requires automated partitioning of the grid. Manually determining which regions must be separated into distinct VE regions is not practical for grids with a large number of cells. Segmenting the domain into different regions can be done using a simple two-step algorithm. We first define an indicator $N_{ij}$ to be 1 if cells $i$ and $j$ are connected with transmissibility below some threshold, or otherwise specified by the user. We define a column-connectivity matrix $A_c$,
\begin{equation}
(A_c)_{ij} =
\begin{cases}
1 \mbox{ if } N_{ij} \land C(i) = C(j)\\
0 \mbox{ otherwise.}
\end{cases}
\end{equation}
Taking the disconnected components of $A_c$ using a depth-first search from the Boost Graph Library package \cite{bgl,matlab-bgl}, will give us all columns of the grid, divided into sections where either missing geometric connections or transmissibility below a defined threshold block flow as the vector $c_c$. We now have a number of coarse blocks that each belong to a single column, with possibly multiple blocks that belong to the same column. Define $G(i)$ to be the column number of block $i$ and let $M(i, j)$ be the number of blocks with column number $j$ that block $i$ is connected to with a non-zero transmissibility. We define the block connectivity matrix,
\begin{equation}
(A_b)_{ij} =
\begin{cases}
1 \mbox{ if } M(i, G(j)) = 1 \\
0 \mbox{ otherwise,}
\end{cases}
\end{equation}
and take the block connected components $c_b$. If we treat the original partition as a linear index into this next connected components, $c_b(c_c)$, we merge any columns that do not transition to more than one column as seen in Figure~\ref{fig:coarsening} and obtain the discrete mapping $V$ required in \eqref{eq:coupled_flux}.

\section{Numerical examples}
For the numerical examples, we will consider two conceptual problems and two corner-point field models. As property modeling is not the focus of the discussion, we will use the properties given in Table~\ref{tbl:props} for \co and brine. All examples include the effect of compressibility, except for the first motivational example. All solvers use the same time-steps. These are generally chosen for the fine-scale reference, where very long time-steps can be challenging when column-segregation must be resolved.

\begin{table}[ht]
	\centering
	\caption{Fluid properties used in the examples. Reference densities are given at 100 bar pressure.}
	\label{tbl:props}
	\begin{tabular}{l|rrr}
		Pseudocomponent & Compressibility & Viscosity & Reference density\\ \hline
		Brine (liquid) & $10^{-7}$ bar$^{-1}$ & 0.80 cP & 1200 kg/m$^3$ \\
		\co (vapor)   & $10^{-4}$ bar$^{-1}$ & 0.06 cP & 760  kg/m$^3$
	\end{tabular}
\end{table}
\subsection{A motivational example: Gravity-dominated displacement}
As the first example, we will consider a simple problem where applying vertical-equilibrium results in considerable reduction in the number of degrees-of-freedom, without any loss of accuracy. We will demonstrate the transition between fine-scale and VE discretizations and consider the effect of relative permeabilities on the simulation results.

We consider a two-dimensional domain with width 10,000 m and a depth of 10 m. The domain has a permeability of 300 mD and uniform porosity of 0.3. One pore-volume of \co is injected at the $x=0$ over 100 years. The boundary at $x=10,000$ m is kept at fixed pressure. The parameters of this problem are within ranges typically considered for \co storage problems, with decades of injection into an aquifer spanning several kilometers.

\begin{figure}[ht]
	\centering
	\subfloat[Fine grid ($100\times100=10,000$ cells)]{
		\includegraphics[width=.32\textwidth]{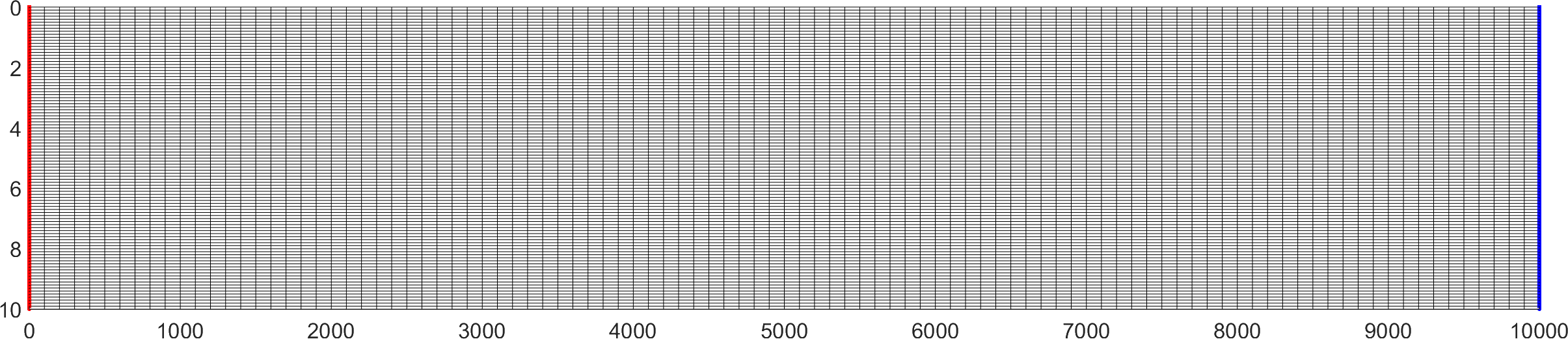}}
	\subfloat[Coarse grid ($100\times1=100$ cells)]{
		\includegraphics[width=.32\textwidth]{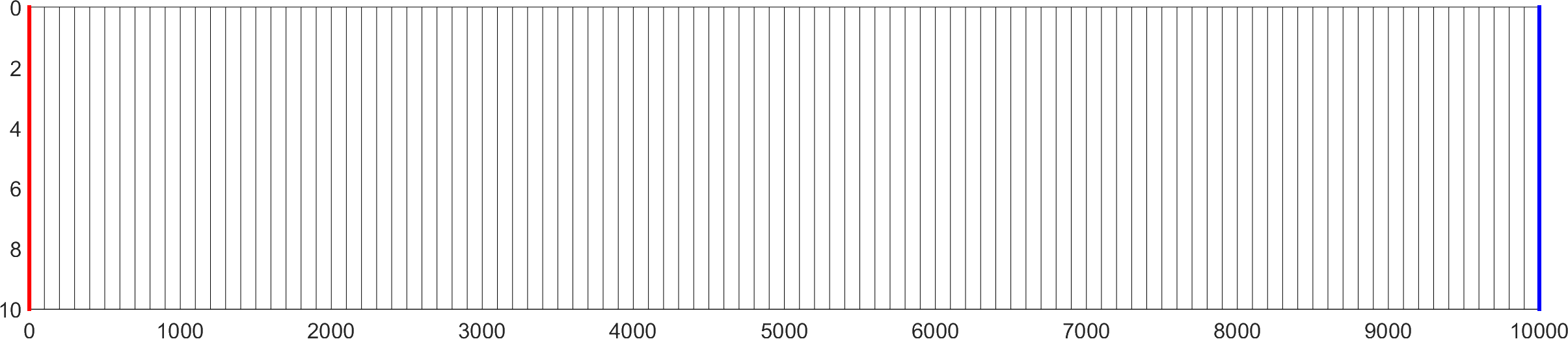}}
	\subfloat[Hybrid grid (3862 cells)]{
		\includegraphics[width=.32\textwidth]{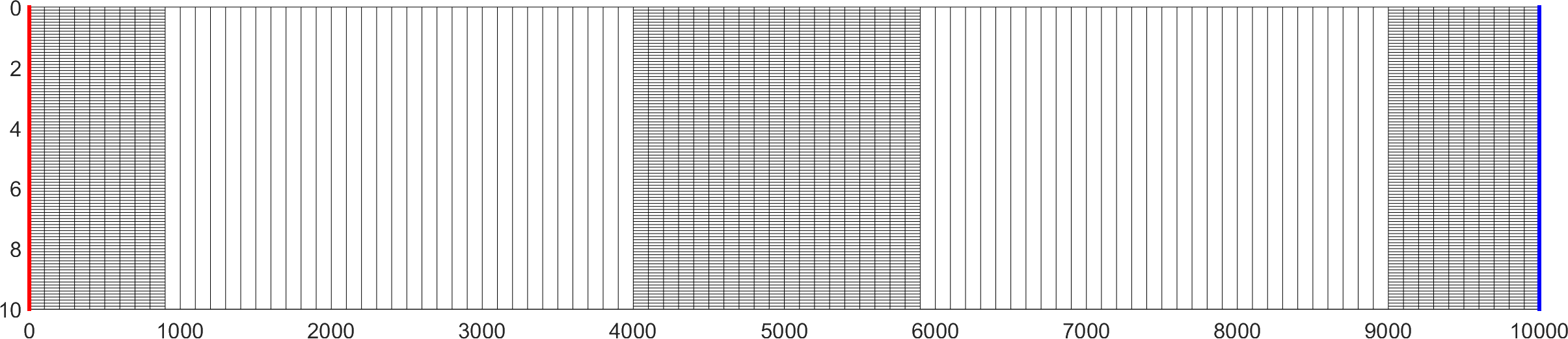}}
	\caption{}
	\label{fig:simple_grids}
\end{figure}

Three choices of grid resolution, shown in Figure~\ref{fig:simple_grids} and four different solvers are considered. For a reference solution, we consider the fine-scale solver, with full vertical resolution of 100 cells in each column. Two different VE-models are considered: One conventional VE-model with a single cell in each column and a hybrid solver with refinement in three regions as shown in Figure~\ref{fig:simple_grids}: The first and last 10\% of the domain uses 100 cells in the vertical, as well as 20\% of the distance centered around the midway point. Finally, we compare with a coarse-scale solver which uses the same resolution as the conventional VE-model.

\begin{figure}[ht]
	\centering
	\subfloat[Linear relative permeabilities]{
		\includegraphics[width=.5\textwidth]{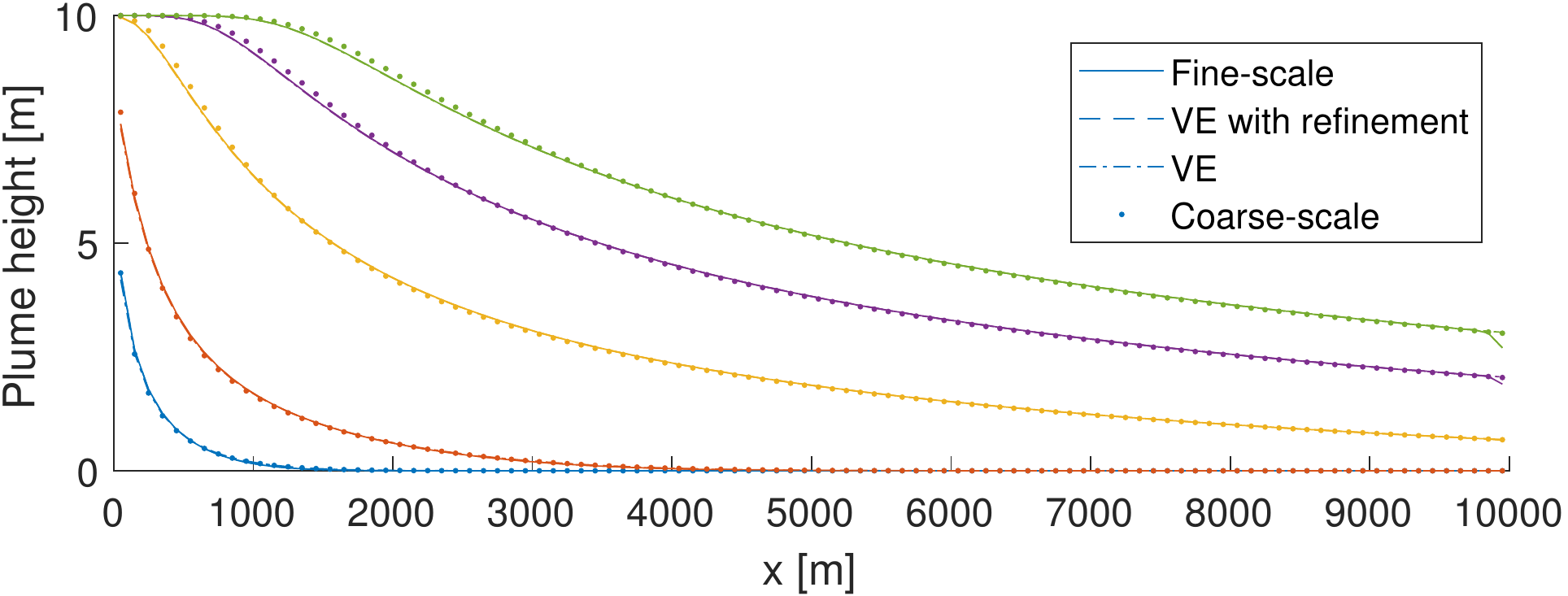}}
	\subfloat[Nonlinear relative permeabilities]{
		\includegraphics[width=.5\textwidth]{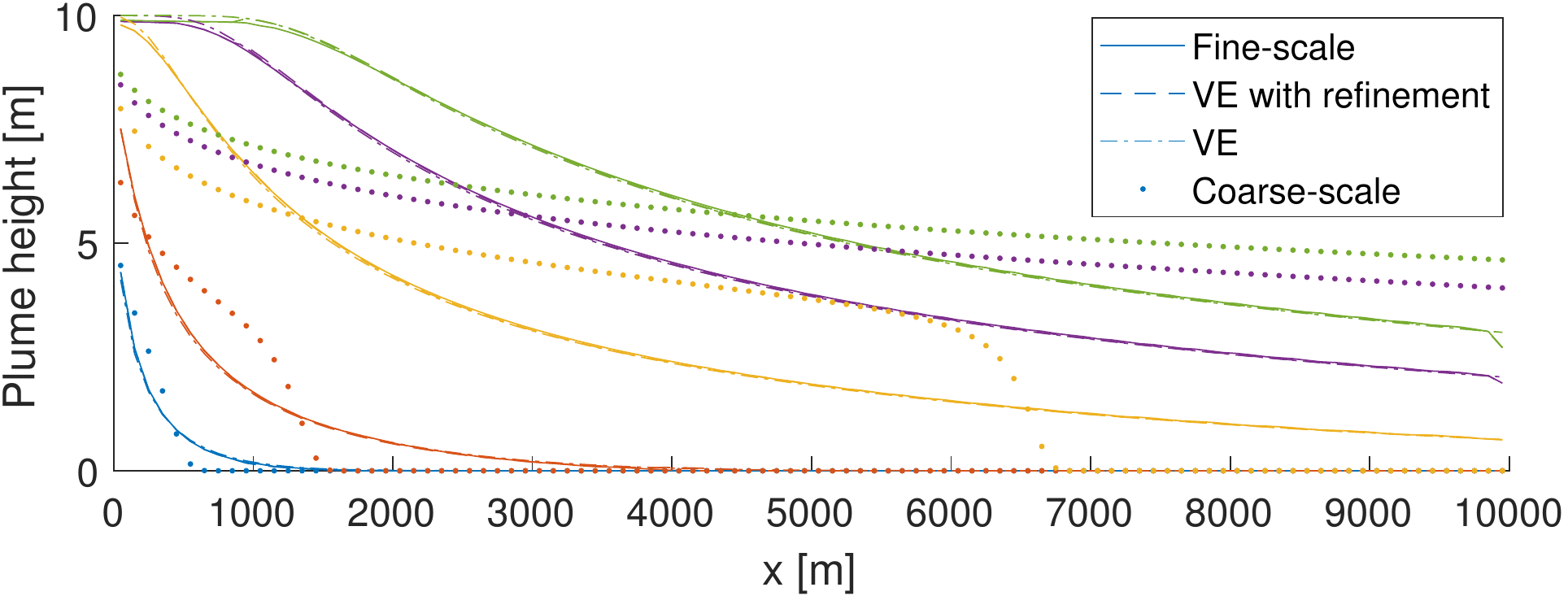}}
	\caption{The gravity-dominated displacement shown with linear and nonlinear relative permeabilities. In problems with local gravity segregation, the choice of permeability curve has negligible impact on the plume profile. Heights are plotted at 0.01, 0.05, 0.3, 0.66 and 1 pore-volumes injected (from left to right).}
	\label{fig:simple_displacement}
\end{figure}

We first consider the case where relative linear permeabilities are considered, i.e. $k_{r\water}(S_\water) = S_\water$ and $k_{r\gas}(S_\gas) = S_\gas$. In this case, the fine-scale relative permeabilities agree with the integrated values used in the VE models and we will expect the four solvers to produce similar results. For our second case, we consider the case of nonlinear relative-permeabilities on the fine-scale: Brook-Corey functions is used for both water and gas with different coefficients, $k_{r\water}(S_\water) = S_\water^3$ and $k_{r\gas}(S_\gas) = S_\gas^2$. We plot the results at several different time-steps corresponding to early, intermediate and late parts of the injection for both cases in Figure~\ref{fig:simple_displacement}.

We observe that all solvers are in general agreement for the case with linear relative permeabilities. Aside from a small coning effect after breakthrough near the end of the domain only present in the models with multiple vertical cells, the solutions are close to identical. The hybrid model produces accurate transitions between the zones and it is not possible to delineate where the transition happens from the height plot. For non-linear permeabilities, we observe that there is not a large deviation between the coarse model without VE and the remainder of the models. Comparing to the linear case, there is not much visible difference in the fine-scale solutions, indicating that linear relative permeabilities are accurate for buoyancy dominated problems. This effect happens because all but one cell in a vertical column will be evaluated with either zero or unit saturations. For this reason, care should be taken when applying relative permeability curves evaluated in one flow-regime and grid resolution to a completely different scale and flow-type.

\FloatBarrier
\subsection{Conceptual model with layered structure}
\begin{figure}[ht]
	\centering
	\subfloat[Fine-scale model (20,000 cells)]{
		\includegraphics[width=.46\textwidth]{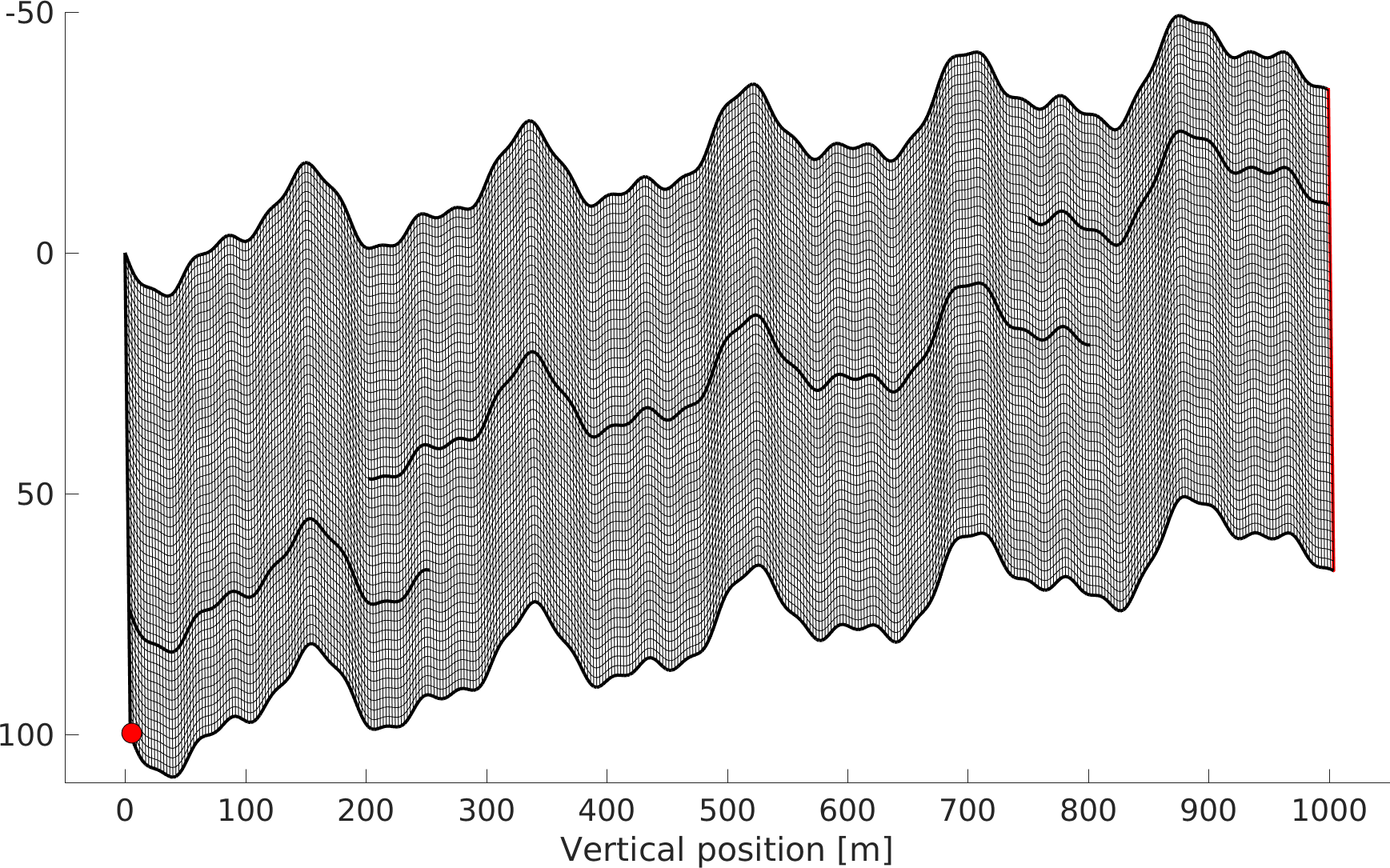}} \hspace{.25cm}
	\subfloat[Hybrid model (1368 cells)]{
		\includegraphics[width=.46\textwidth]{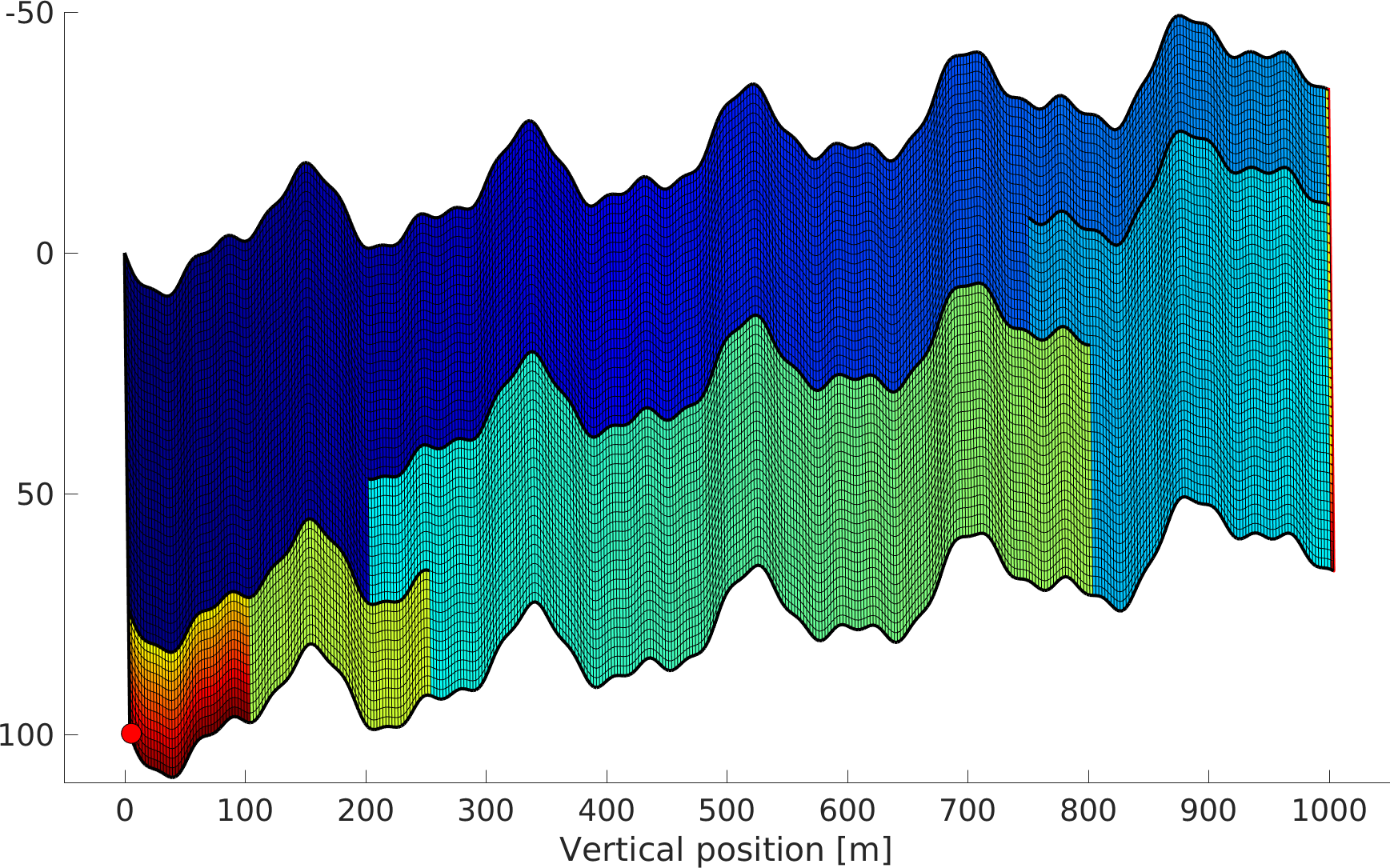}} \hspace{.25cm}\\
	\caption{A synthetic 2D model with CO$_2$ injection in the lower left corner and an open boundary at the right-most part of the domain (left). The model is coarsened into multiple VE-zones (right), including a fine-scale near-well region shown in red.}
	\label{fig:simple_setup}
\end{figure}

We will consider a conceptual synthetic model with several impermeable layers where conventional VE does not produce reliable results. The aquifer is shown in Figure~\ref{fig:simple_setup} and contains $400\times50$ cells with a uniform permeability of 100 mD. The model is tilted by 2.5 degrees, resulting in upslope flow over a top surface with undulations with three different periods, giving a complex trapping structure for the lighter CO$_2$. Three impermeable layers are introduced by applying a transmissibility multiplier of zero. These layers will divert the flow from the injection site and significantly impact the final CO$_2$ distribution. The schedule is set to an injection of 20\% pore-volume CO$_2$ over a period of 30 years, followed by a migration period of 970 years.

\begin{figure}[ht]
	\centering
	\subfloat[End of injection, fine-scale]{
		\includegraphics[width=.32\textwidth]{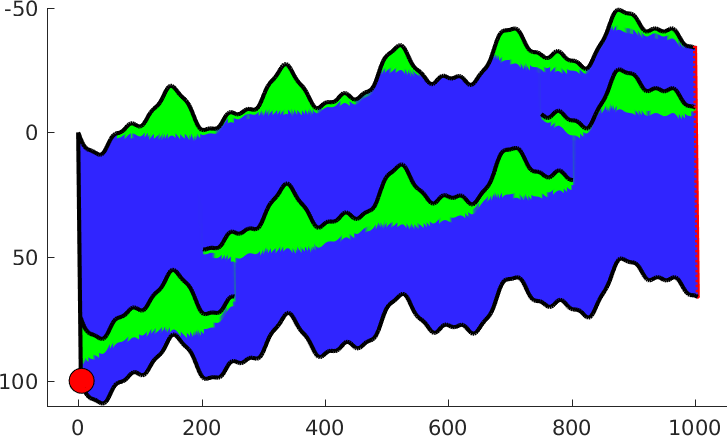}}
	\subfloat[End of injection, hybrid vertical-equilibrium]{
		\includegraphics[width=.32\textwidth]{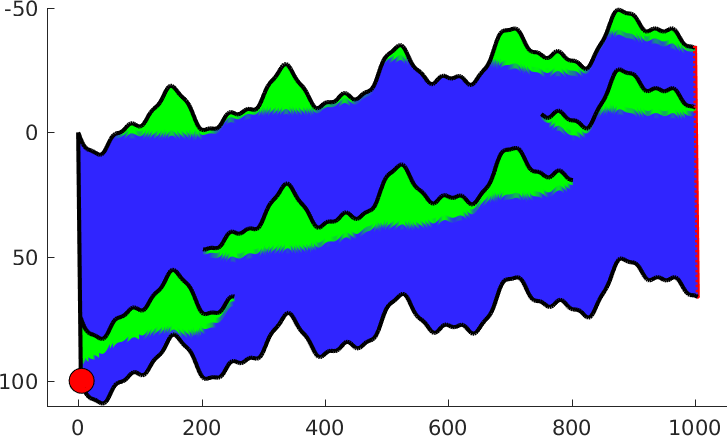}}
	\subfloat[End of injection, only vertical-equilibrium]{
		\includegraphics[width=.32\textwidth]{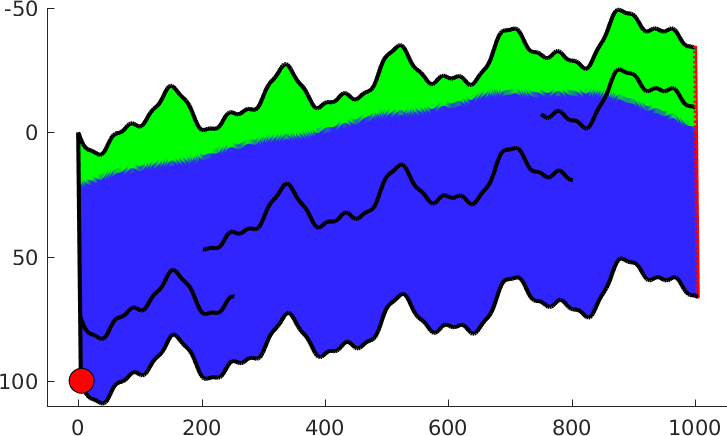}}\\
	\subfloat[End of migration, fine-scale]{
		\includegraphics[width=.32\textwidth]{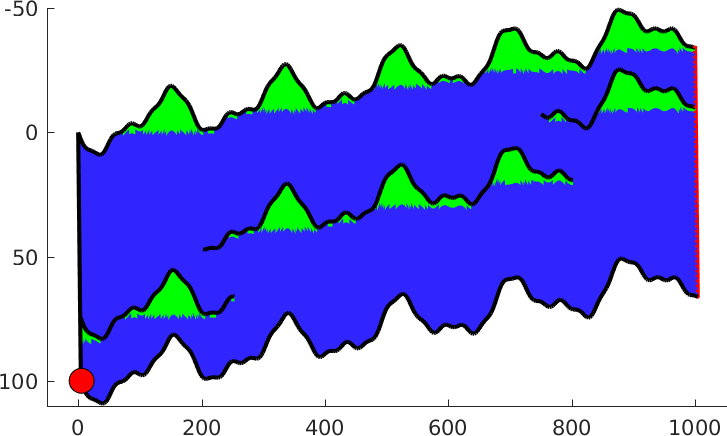}}
	\subfloat[End of migration, hybrid vertical-equilibrium]{
		\includegraphics[width=.32\textwidth]{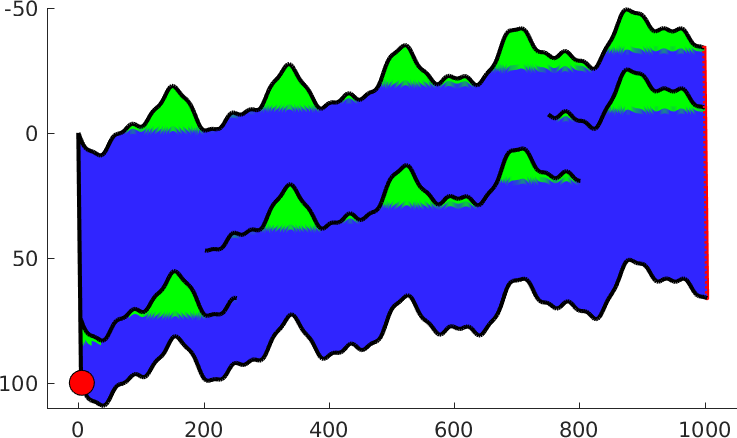}}
	\subfloat[End of migration, only vertical-equilibrium]{
		\includegraphics[width=.32\textwidth]{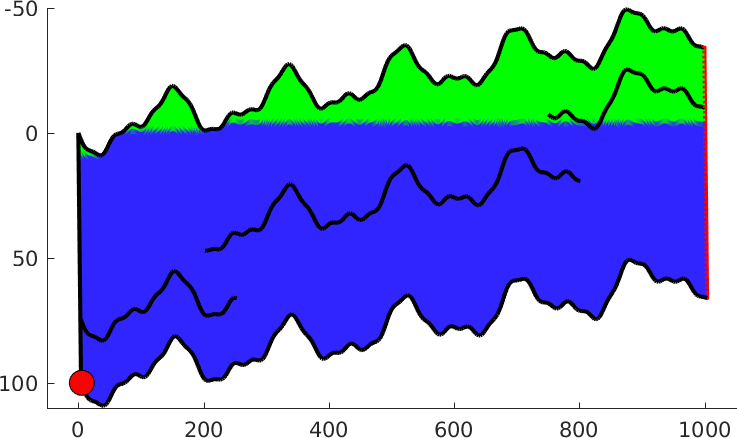}}
	\caption{Saturation plots for the conceptual model with impermeable layers. CO$_2$ injection in the lower left corner migrates towards the open boundary at the right side of the domain. Conventional VE models assume vertical equilibrium throughout the vertical direction and are unable to predict the final trapping structure.}
	\label{fig:simple_sloped}
\end{figure}

Three different solvers are used: The fine-scale solver using a conventional finite-volume discretization, a vertical-equilibrium solver without layers and a hybrid solver with VE regions between the impermeable layers, as well as fine-scale resolution in the near-well region. For this conceptual example, we will not consider the layered solver without near-well refinement. All solvers are fully-implicit, allowing for large time-steps even with significant compressibility. The results can be viewed in Figure~\ref{fig:simple_sloped}, where the CO$_2$ saturation for all three solvers are plotted at the end of the injection and migration periods. The impermeable layers significantly impact the migration of CO$_2$, resulting in the VE model with a single cell per column in the grid over-estimating the time until the gas reaches the caprock. The final trapping structure is also different. The hybrid model has excellent agreement with the the fine-scale model. The transition between the fine-scale and the vertical equilibrium is accurately captured, as is  the transition between different regions under the VE regime. The same case was also simulated with quadratic and cubic relative permeability curves for the fine-scale model, without any significant difference in the results -- as this is in agreement with the discussion in Example 1, we do not report these results here.

\FloatBarrier
\subsection{Refined subset of Sleipner model}
The Sleipner field in the North Sea of Norway is a site for \co re-injection, where 16.5 million tonnes of \co has been injected into the Utsira formation over the 20 years of operational history \cite{Cavanagh2015}. Since it was the first of the large scale injection for storage it has been used as benchmark on simulation framework. In addition a model of the upper layer has been released to facilitate research on realistic data \cite{Singh2010,Sleipner:benchmark}. The upper layer is of particular interest, since this is the region the plume is visible on seismic data.

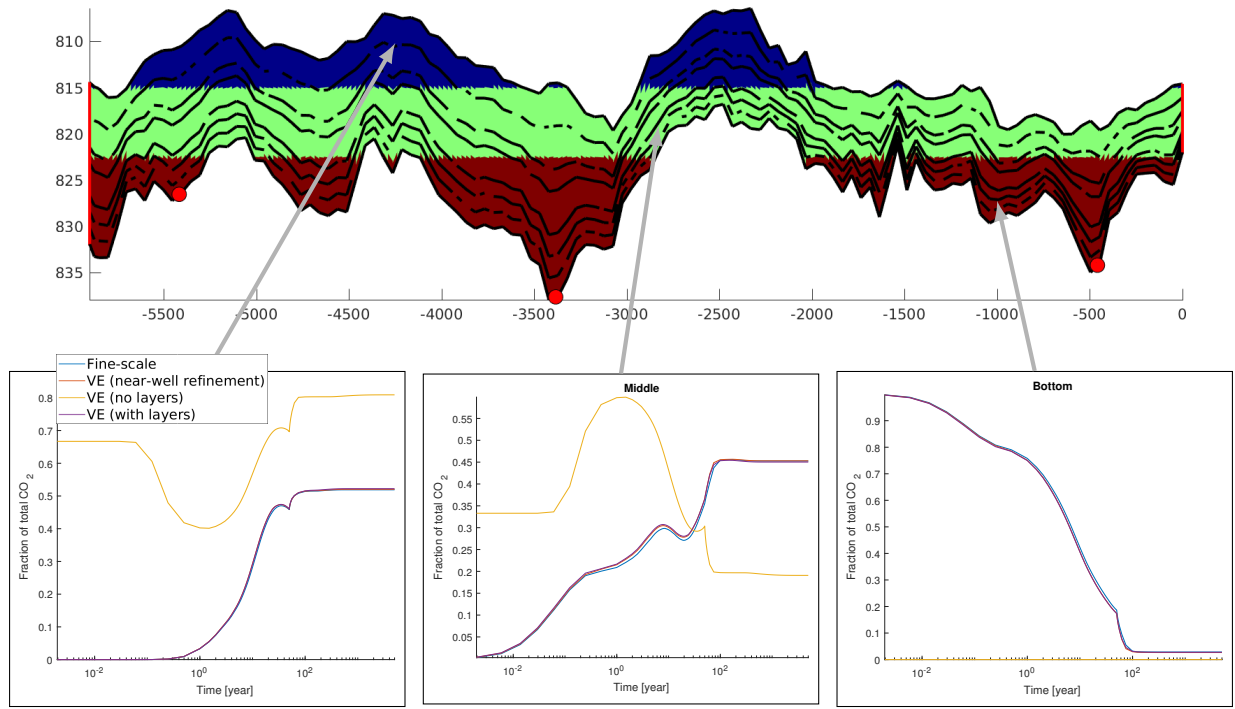
\begin{figure}[htbp]
	\centering
	\begin{tikzpicture}
	\node[anchor=center] at (8,3) {%
		\includegraphics[width=15cm]{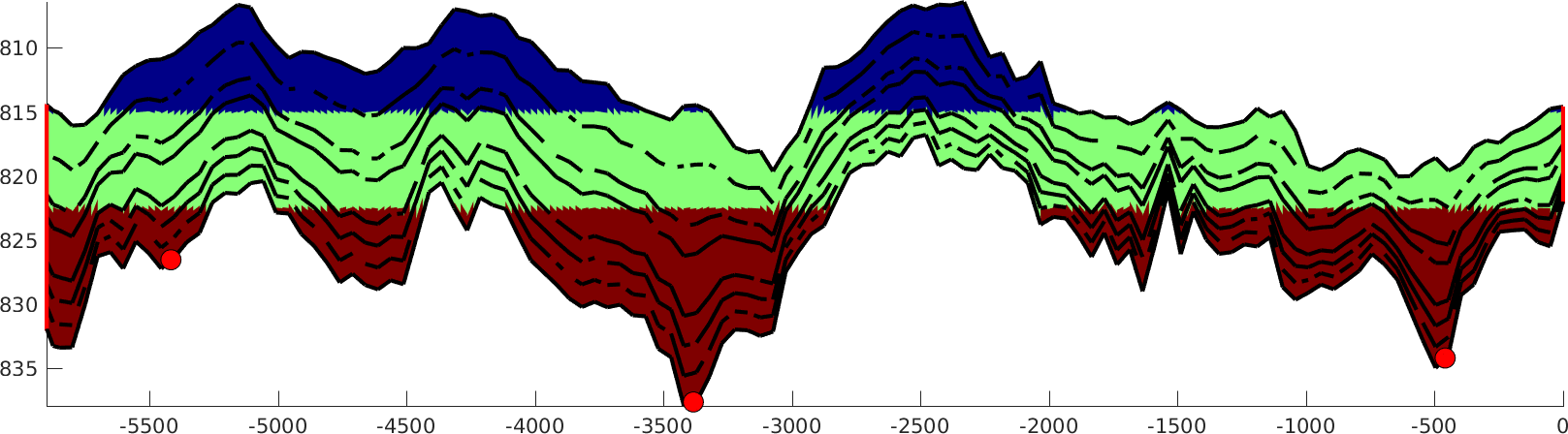}};
	\node[anchor=center,draw=black] at (2.5,-2) (s1) {%
		\includegraphics[width=5.0cm]{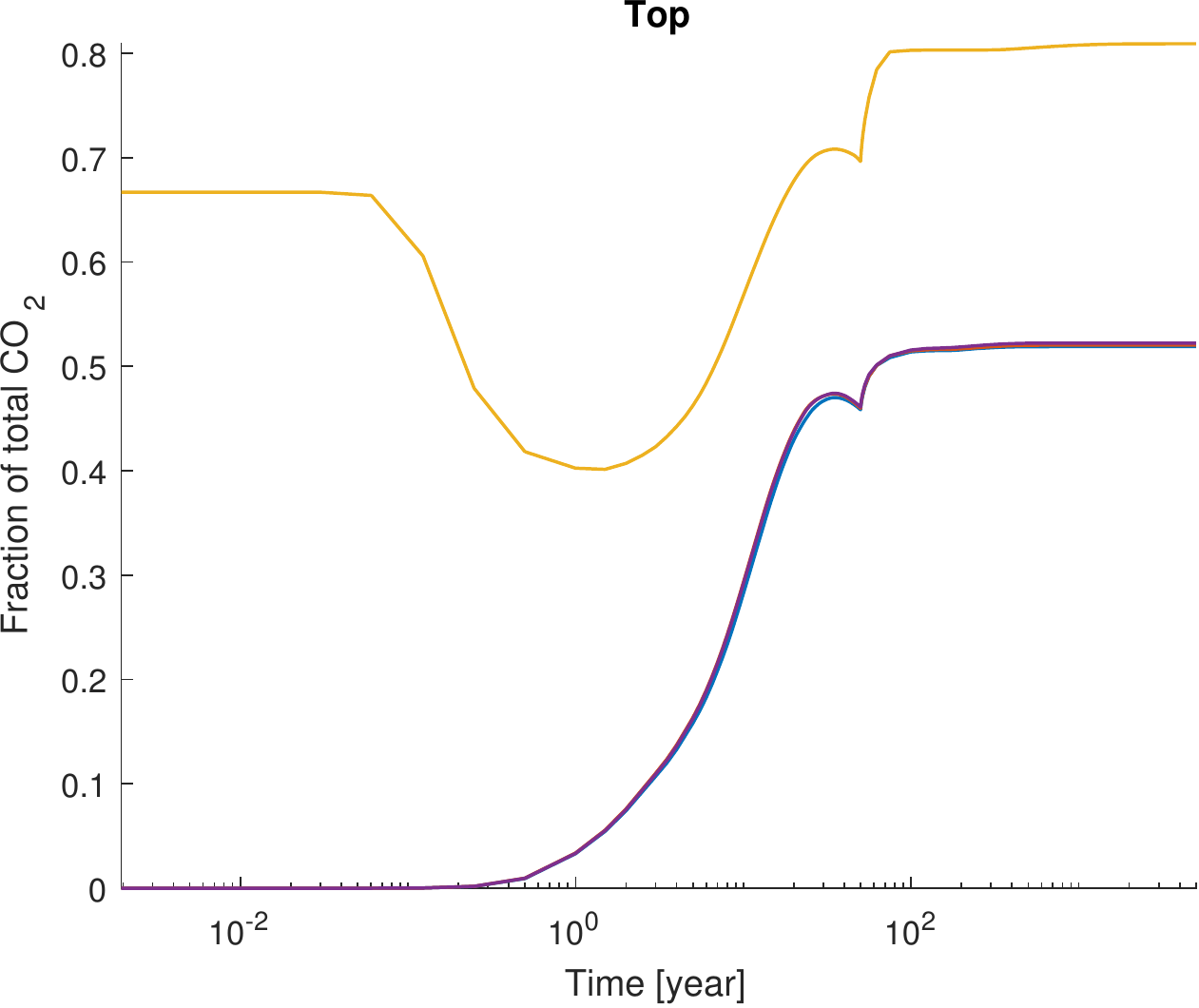}};
	\node[anchor=center,draw=black] at (8, -2) (s2) {%
		\includegraphics[width=5.0cm]{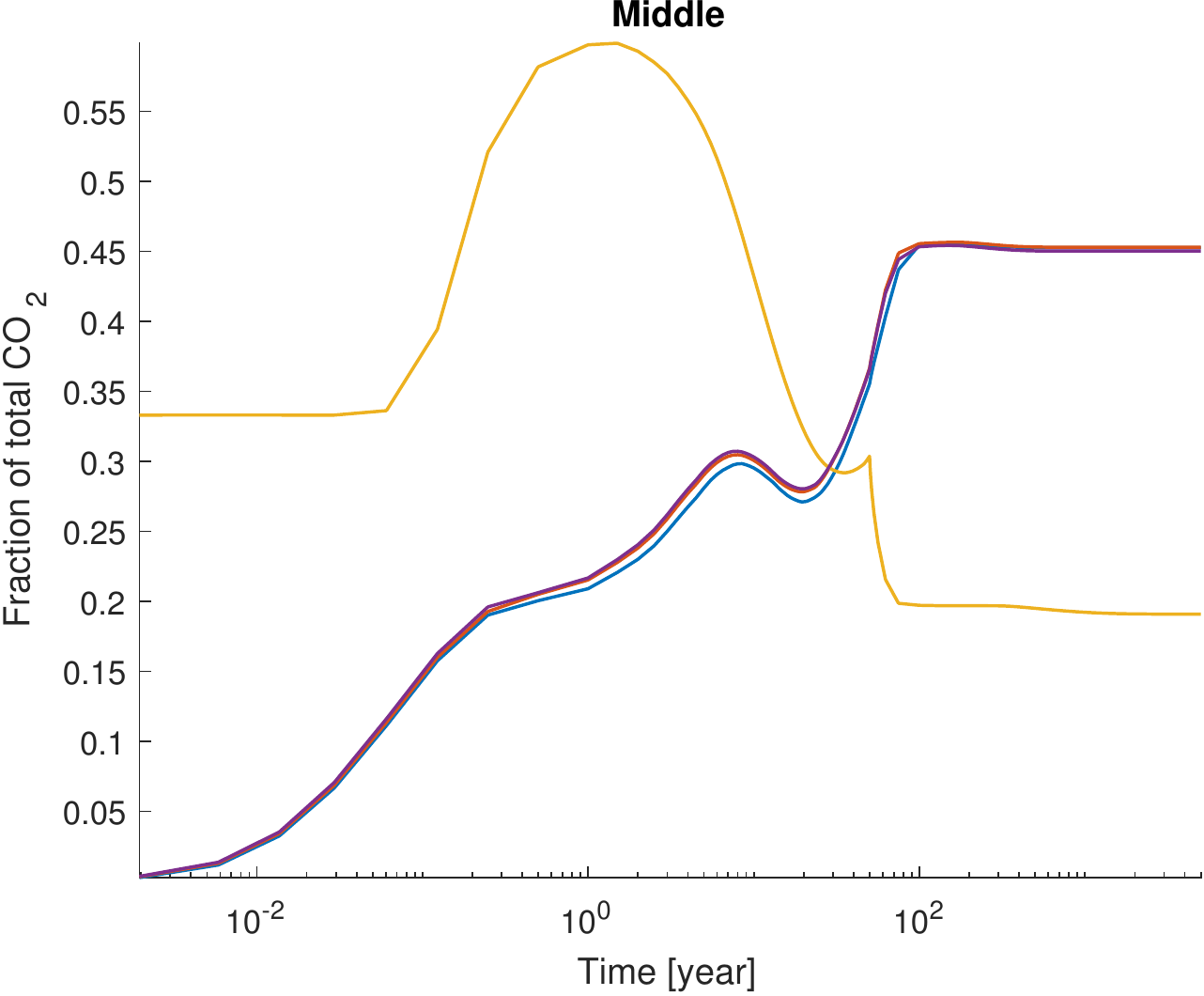}};
	\node[anchor=center,draw=black] at (13.5,-2) (s3) {%
		\includegraphics[width=5.0cm]{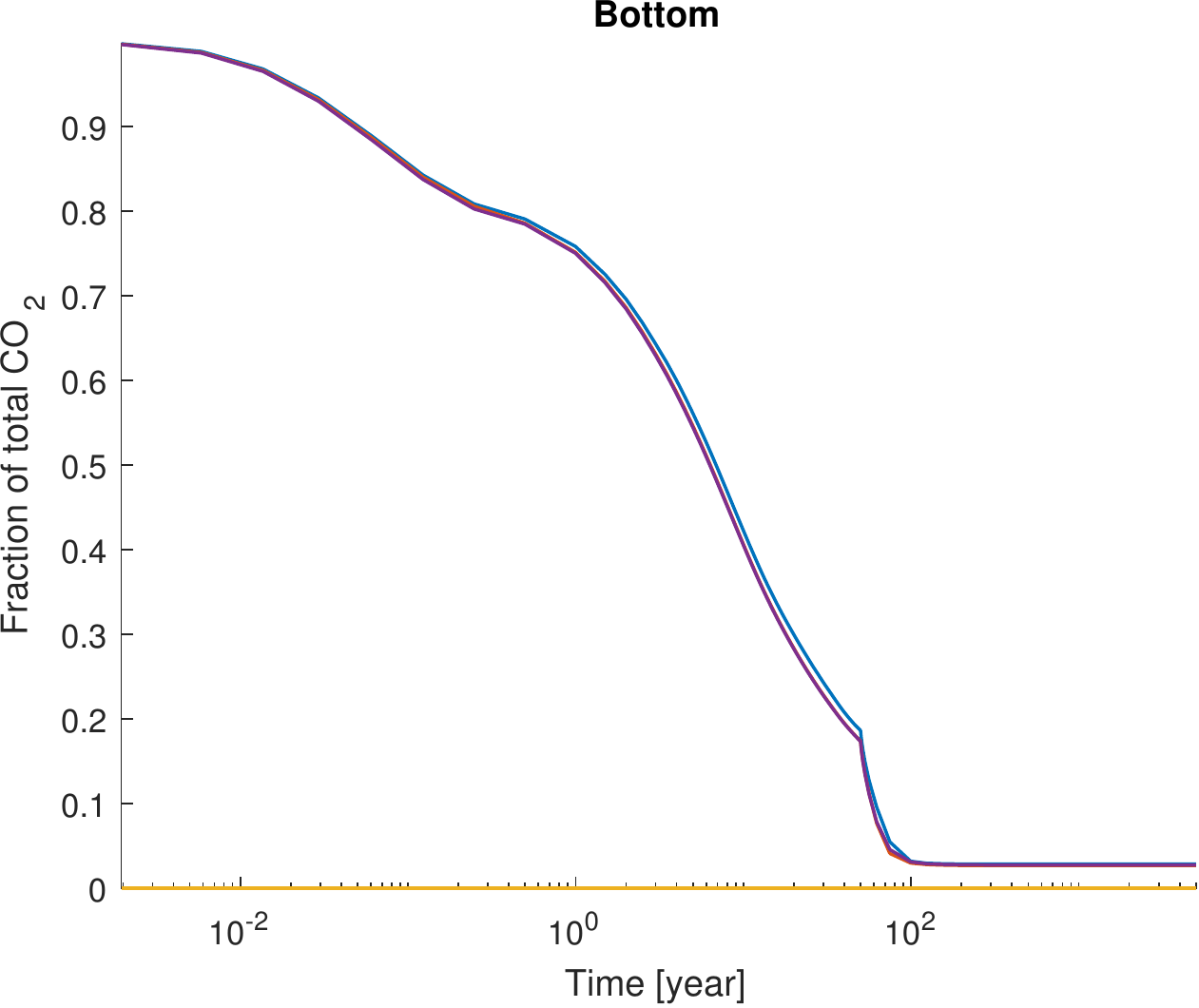}};
	\draw[->,black!30,  ultra thick,-latex] (s1.north) -- ( 5, 4.6);
	\draw[->,black!30,  ultra thick,-latex] (s2.north) -- ( 8.5,3.5);
	\draw[->,black!30,  ultra thick,-latex] (s3.north) -- (13 ,2.5);
	\tiny
	\node[anchor=south east,red]   at (s1.north) {};
	\node[anchor=south east,black] at (s2.south east) {};
	\node[anchor=north east,blue]  at (s3.north east) {};
	\node[anchor=center] at (1.9,0) {%
		\includegraphics[width=2.8cm]{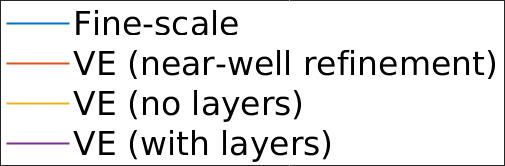}};
	\end{tikzpicture}
	\caption{The setup of the Sleipner subset case. A number of impermeable layers are added to the model, with a stochastic distribution of open segments. The model is divided in to three parts for purposes of analyzing the migration behavior in the different discretized models.}
	\label{fig:sleipner_migration}
\end{figure}

In this example, we will study a complex layered structure with a large number of impermeable barriers. In order to assess the accuracy of the hybrid VE solver, we will use a highly refined two-dimensional subset of the IEAGHG benchmark mode \cite{Singh2010,Sleipner:benchmark}. The grid has logical dimensions of $238\times215$, resulting in 40,460 cells after inactive or pinched cells has been removed. We introduce a layering structure inspired by Figure 2 in \cite{sleipner_basin} where intra-formational shales delay plume migration to the caprock. Six impermeable layers are added, with 25\% of the faces in each layer are left open to flow, representing fractures or eroded shale layers where gas can migrate through. The distribution of open faces in each layer is uniformly random, resulting in a highly complex structure of different flow compartments as shown in the top part of Figure~\ref{fig:sleipner_migration}.

We use a synthetic well configuration where three injection sites at the bottom of the formation inject a total of 0.75 pore-volumes over 50 years, followed by 4950 years of migration. While this is a fairly large amount in terms of total pore-volume, it only corresponds to 5.7 megatonnes of \co due to the thin width of the grid. The formation is open at $x = \min(x)$ and $x = \max(x)$ with hydrostatic boundary conditions.

\begin{figure}[htbp]
	\centering
	\subfloat[End of injection, fine-scale]{
		\includegraphics[width=.45\textwidth]{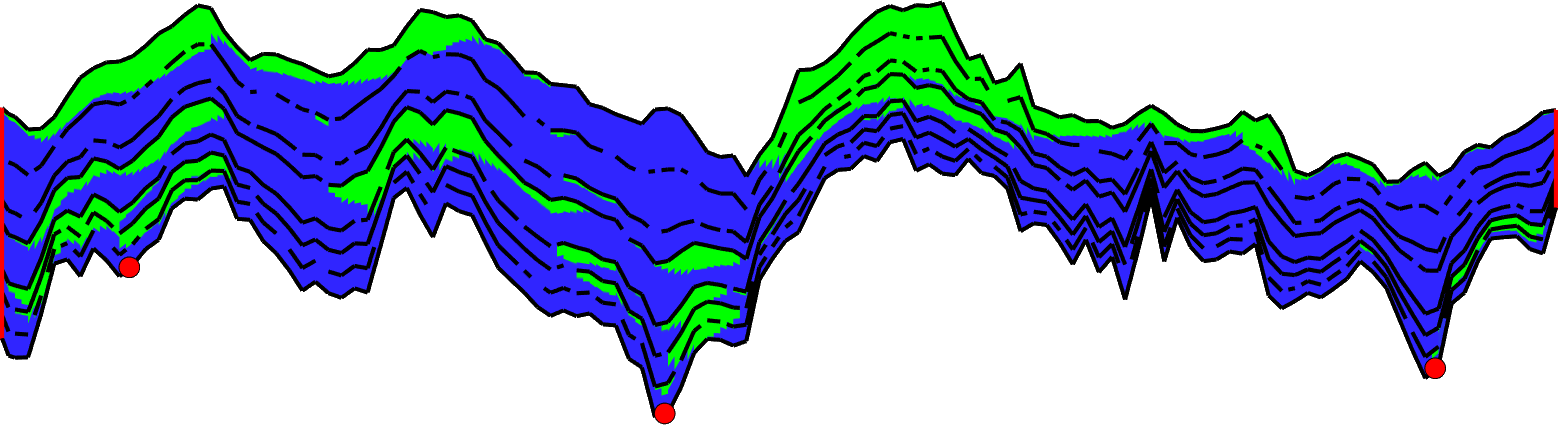}} \hspace{.25cm}
	\subfloat[End of migration, fine-scale]{
		\includegraphics[width=.45\textwidth]{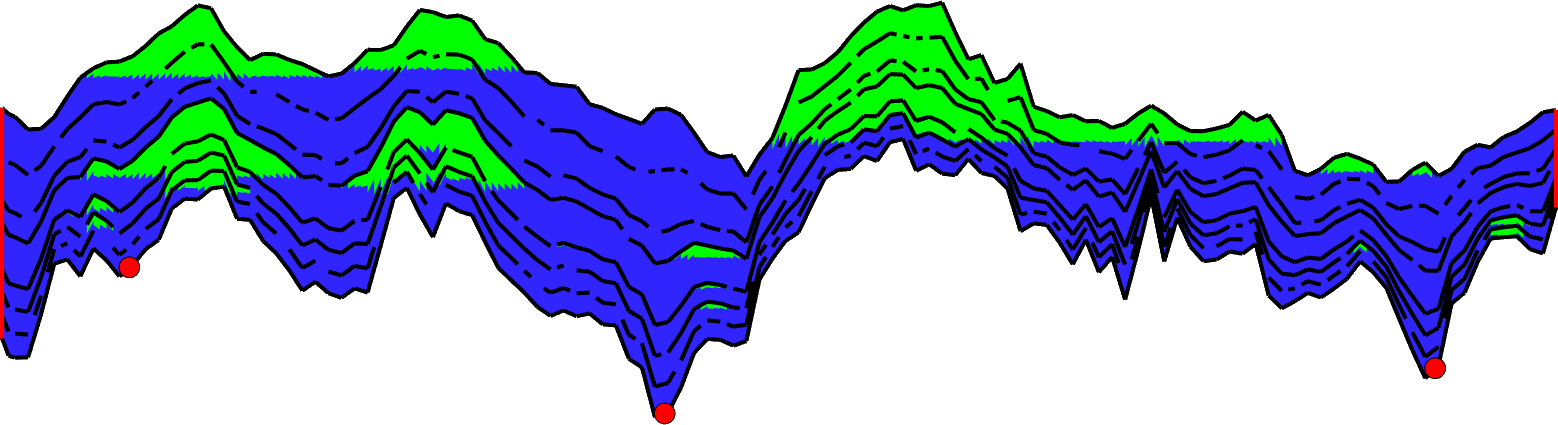}} \hspace{.25cm}\\
	\subfloat[End of injection, hybrid]{
		\includegraphics[width=.45\textwidth]{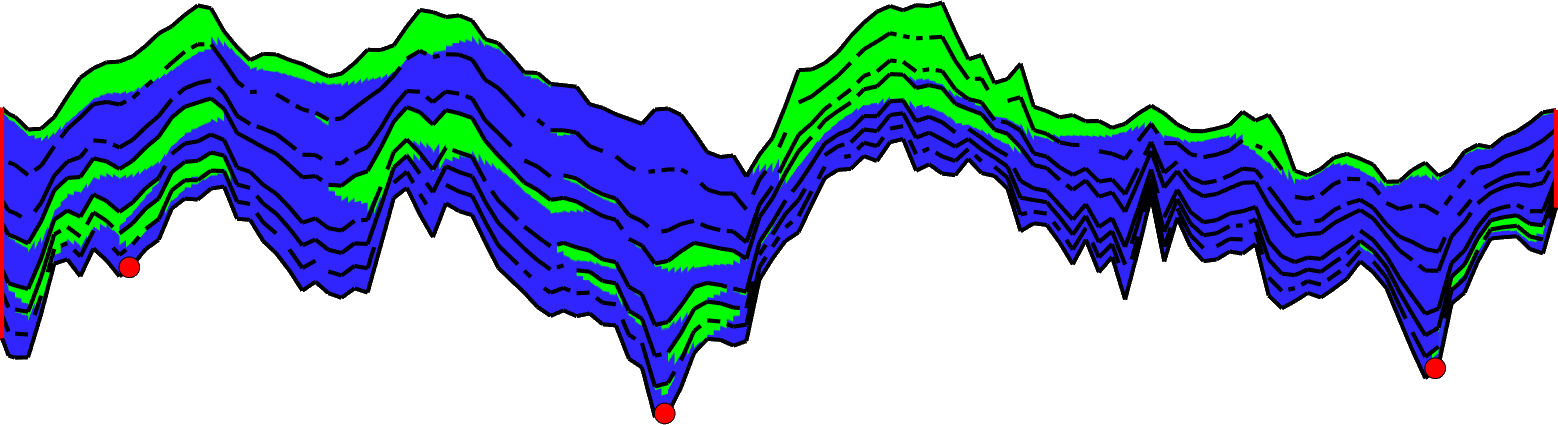}} \hspace{.25cm}
	\subfloat[End of migration, hybrid]{
		\includegraphics[width=.45\textwidth]{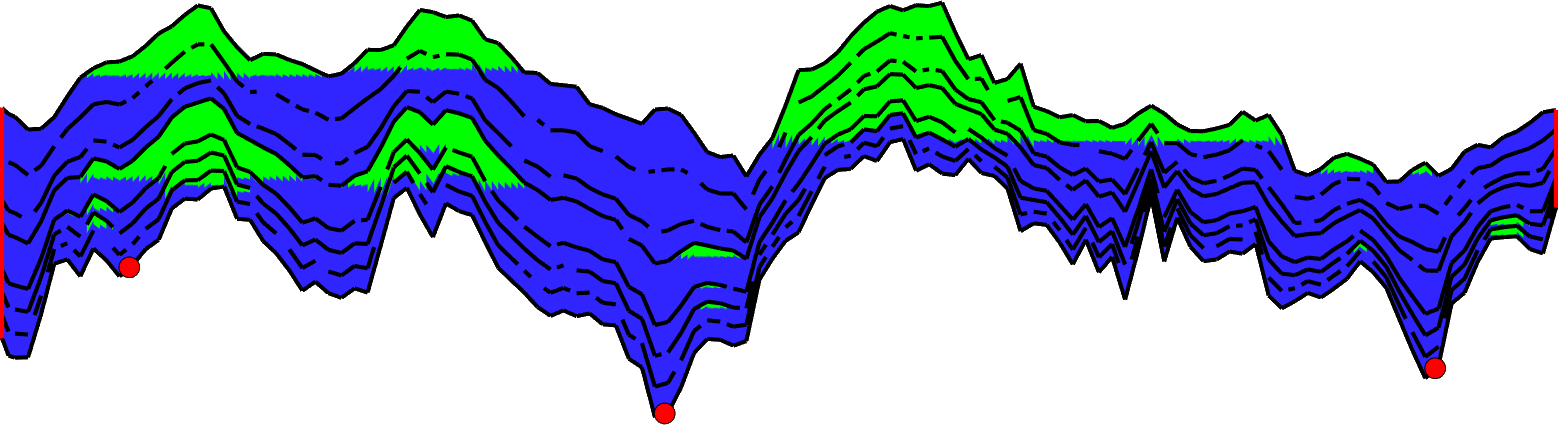}} \hspace{.25cm}\\
	\subfloat[End of injection, vertical-equilibrium]{
		\includegraphics[width=.45\textwidth]{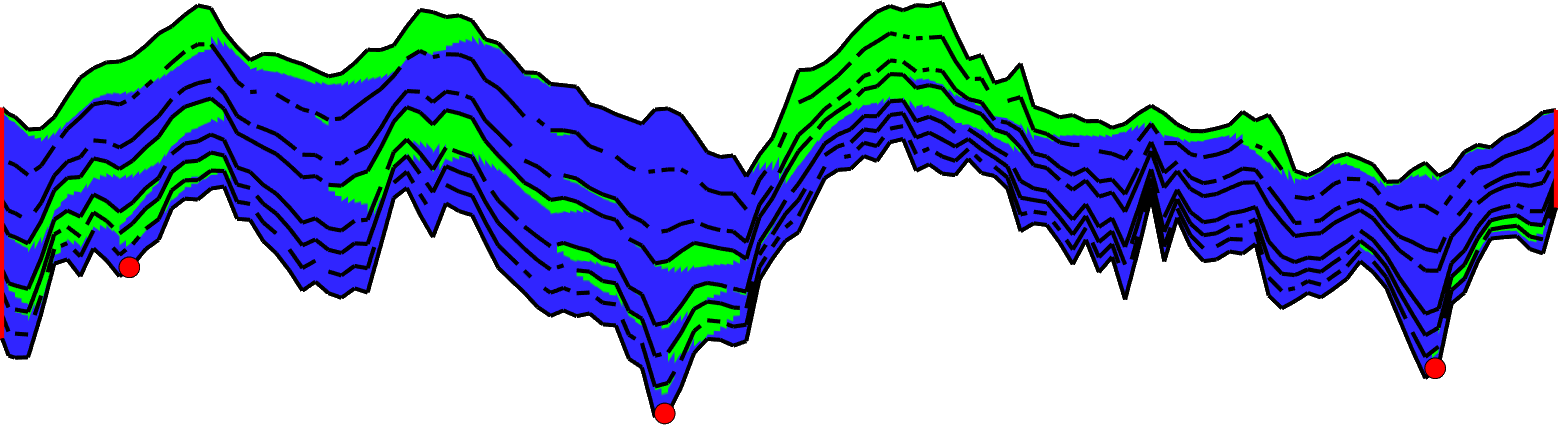}} \hspace{.25cm}
	\subfloat[End of migration, vertical-equilibrium]{
		\includegraphics[width=.45\textwidth]{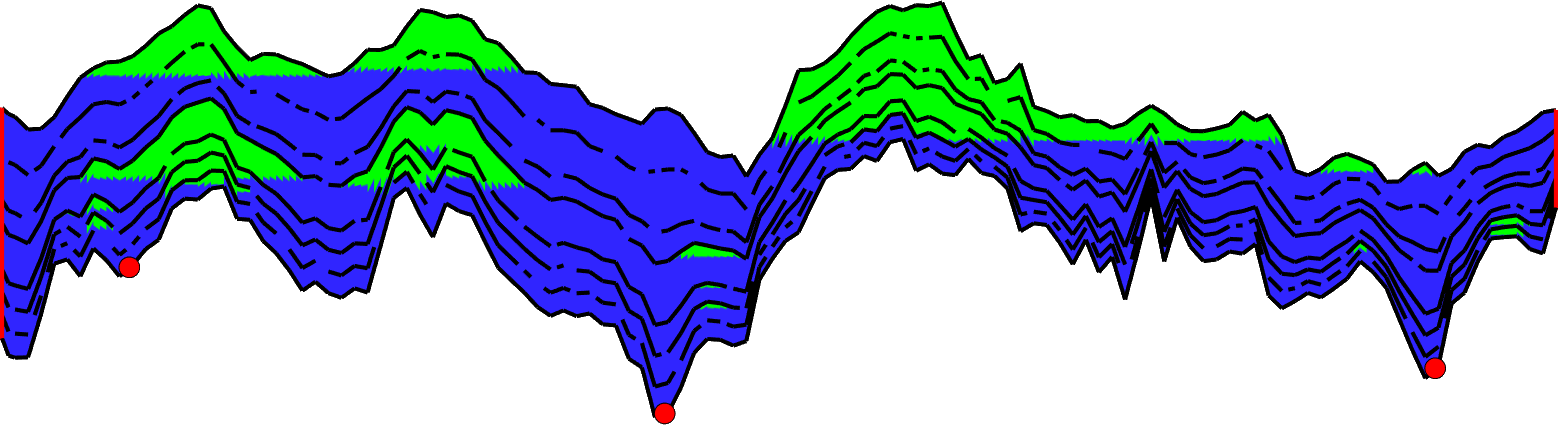}} \hspace{.25cm}\\
	\subfloat[End of injection, vertical-equilibrium (without layers)]{
		\includegraphics[width=.45\textwidth]{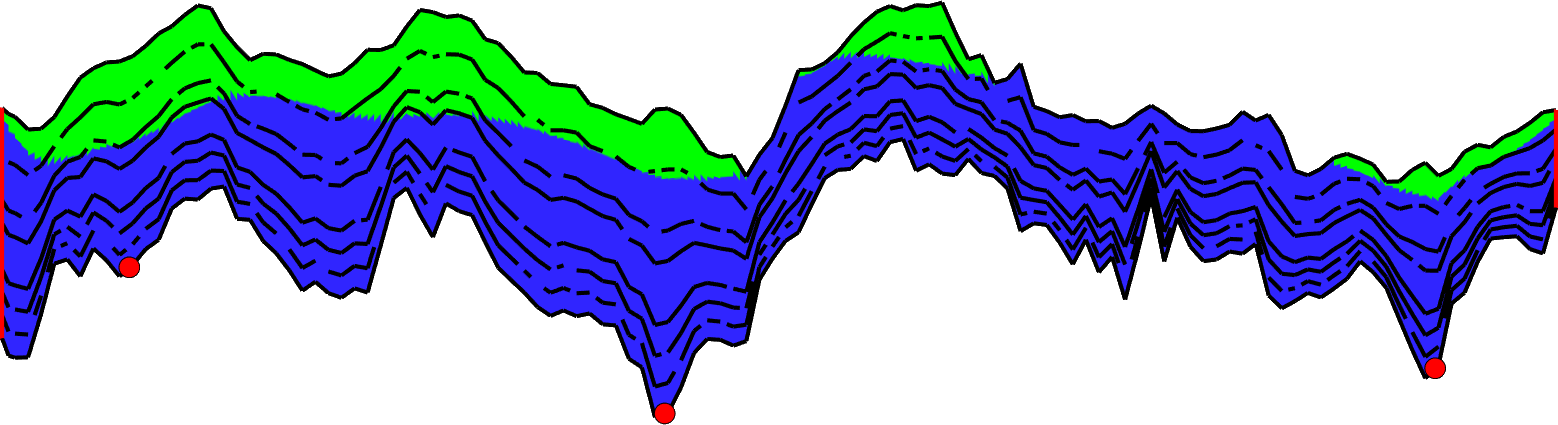}} \hspace{.25cm}
	\subfloat[End of migration, vertical-equilibrium (without layers)]{
		\includegraphics[width=.45\textwidth]{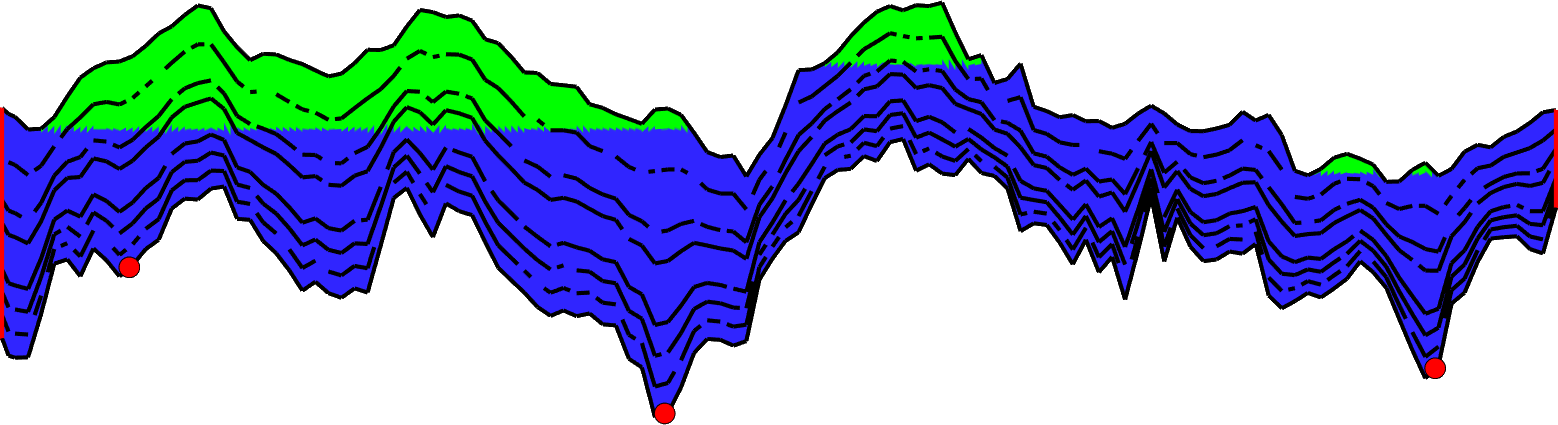}} \hspace{.25cm}\\

	\caption{Reconstructed \co saturation for the four different solvers for the same problem. The left column represents the end of the injection period, while the right column is the stationary solution after 1000 years of subsequent migration. }
	\label{fig:sleipner_sat}
\end{figure}

\begin{figure}[htbp]
	\centering
	\subfloat[BHP, injector 1]{
		\includegraphics[width=.33\textwidth]{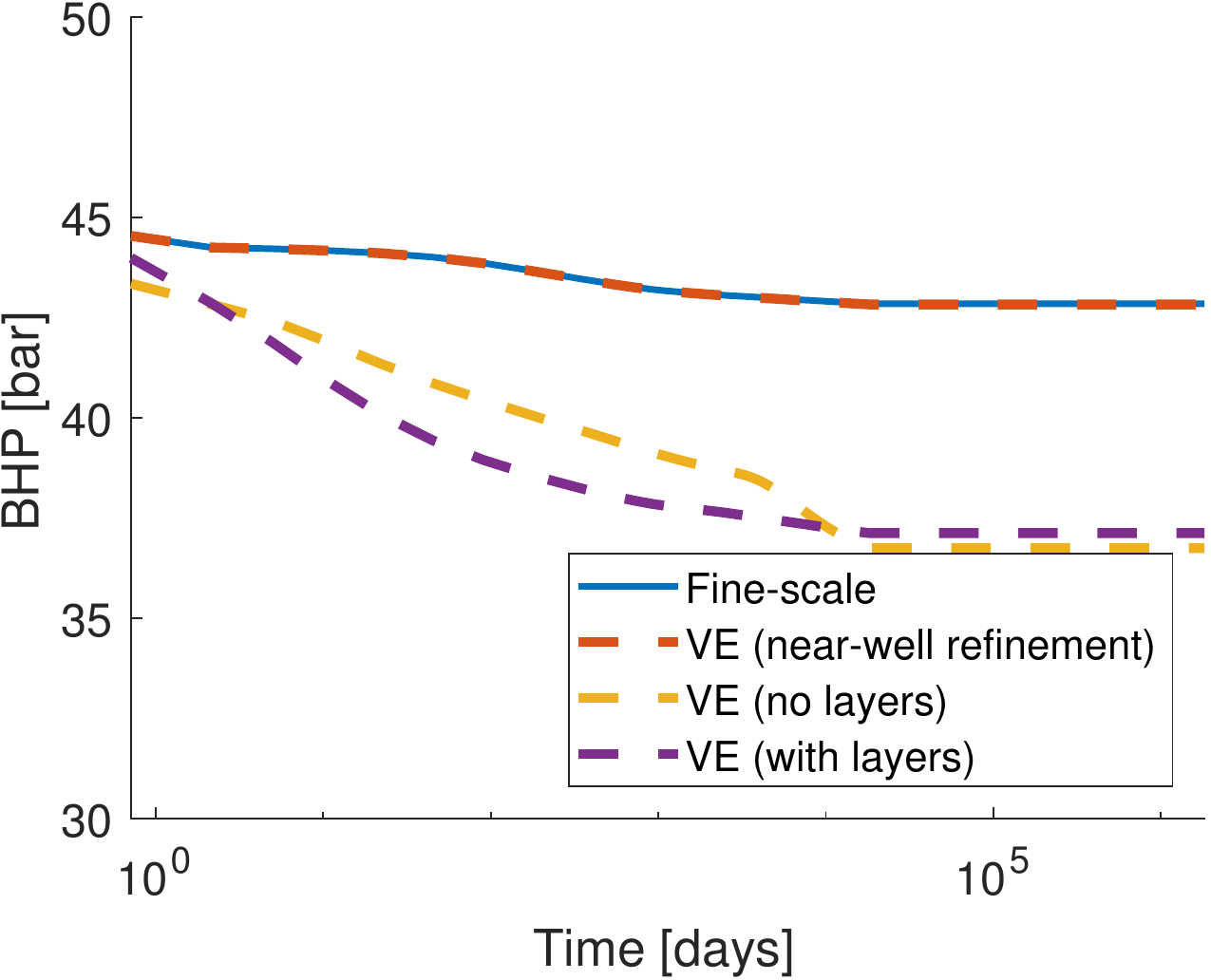}}
	\subfloat[BHP, injector 2]{
		\includegraphics[width=.33\textwidth]{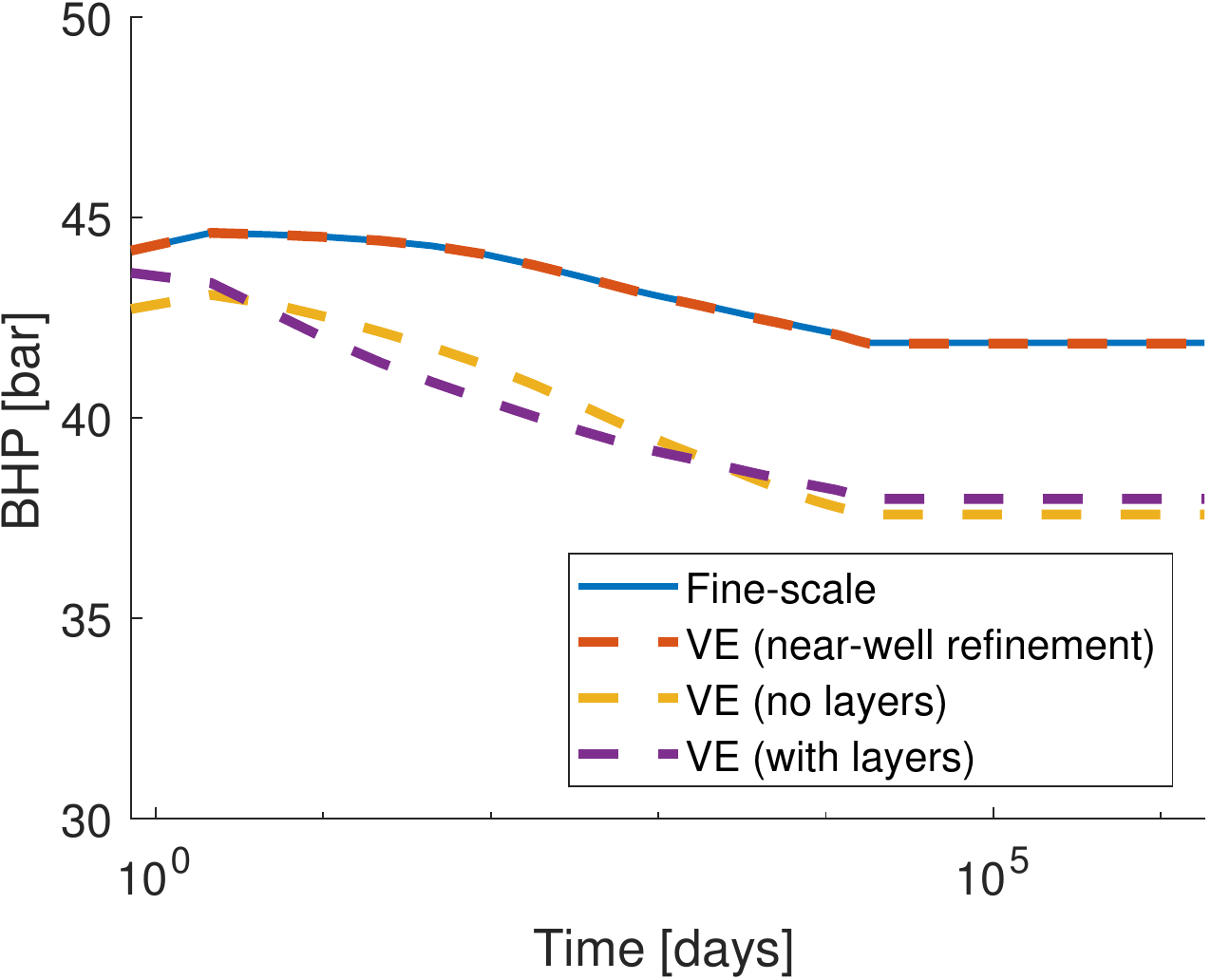}}
	\subfloat[BHP, injector 3]{
		\includegraphics[width=.33\textwidth]{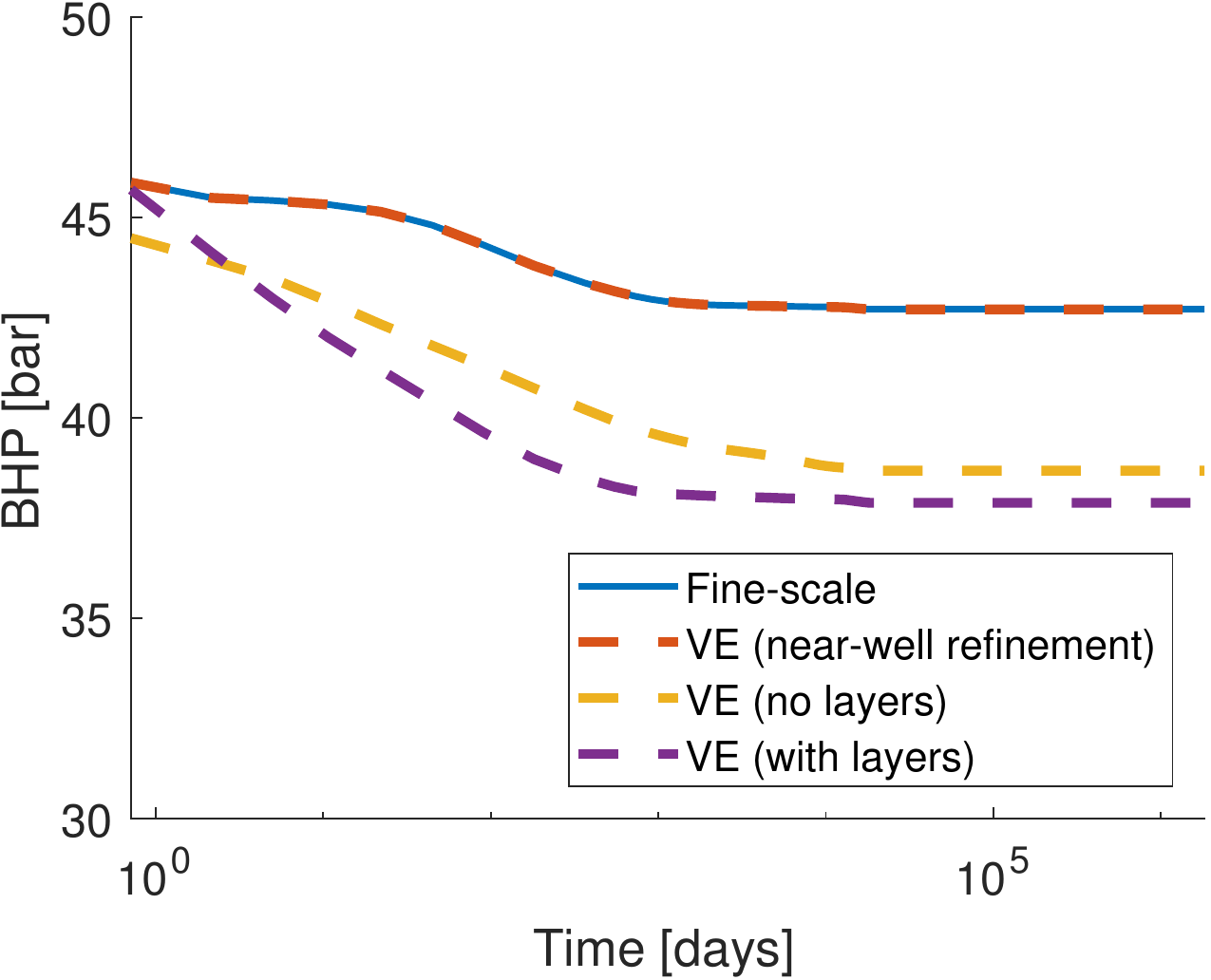}}
	\caption{Bottom hole pressures for the injectors in the Sleipner example during the injection period}
	\label{fig:sleipner_bhp}
\end{figure}

The gas saturation at the end of the injection and migration periods is shown in Figure~\ref{fig:sleipner_sat}. We observe that all solvers produce sharp, horizontal interfaces at equilibrium, but the conventional VE model (700 cells) predicts almost no trapped \co in the right part of the model. The hybrid model with near-well refinement (2700 cells) and the vertical equilibrium with layers included (1776 cells) both accurately predict the saturation distribution at the end of the migration and injection periods. Even insignificant traps of \co is predicted consistently across the fine-scale and the two multiresolution VE-models. In terms of the migration behavior over time, we observe from Figure~\ref{fig:sleipner_migration} that the relative ranking of the four models persist, with agreement between all models which include layers.  If we consider the injector BHP, shown in Figure~\ref{fig:sleipner_bhp}, we see that the hybrid model with fine-scale cells in the near-well region agrees with the fine-scale model in terms of pressure buildup. The models without fine-scale detail does not match the BHP, although a dedicated upscaling procedure for the wells themselves may improve this result. For a real formation introducing specific treatment for the wells may be less practical than including the correct behavior using the fine grid around the well. One would expect that the near-well is relatively stationary at later times and thus will not contribute to the numerical complexity of solving the nonlinear system.

The number of nonlinear iterations used by the different solvers is reported in Figure~\ref{fig:iterations}. We observe that the VE solvers with and without layers use approximately the same low number of iterations when compared to the fine-scale. The solver with near-well refinement uses more iterations, as the segregation process near the well must be resolved by the Newton-Rapshon solver. Relative speed-up by applying VE varies from 548 to 52, depending on the solver used.

\FloatBarrier
\subsection{Layered Utsira model}
\begin{figure}[htbp]
  \centering
  \begin{tikzpicture}
    \node[anchor=center] at (5.5,3) {%
      \includegraphics[width=15cm]{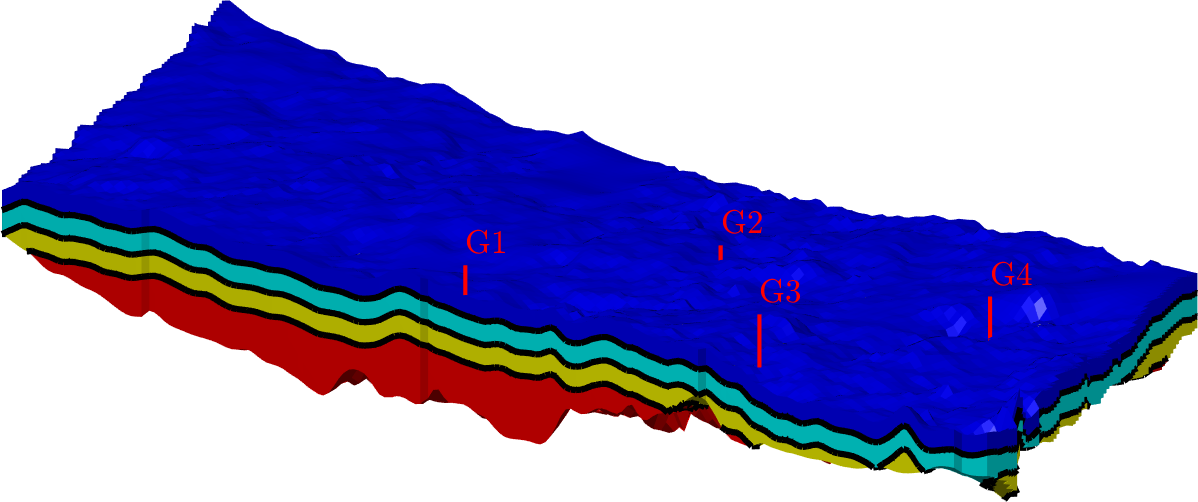}};
    \node[anchor=center,draw=black] at (2.5,9.5) (s1) {%
      \includegraphics[width=7.0cm]{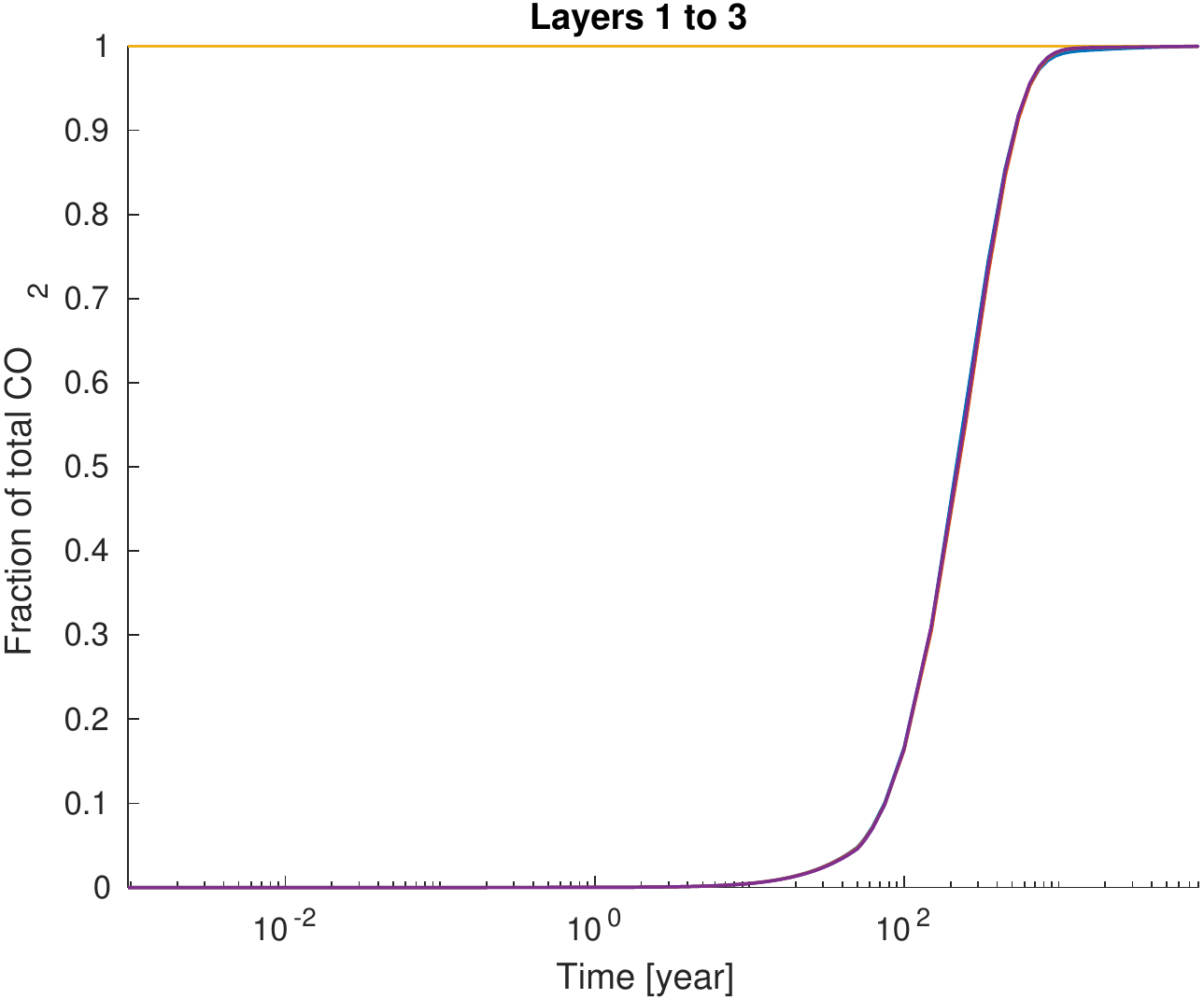}};
    \node[anchor=center,draw=black] at (10, 9.5) (s2) {%
      \includegraphics[width=7.0cm]{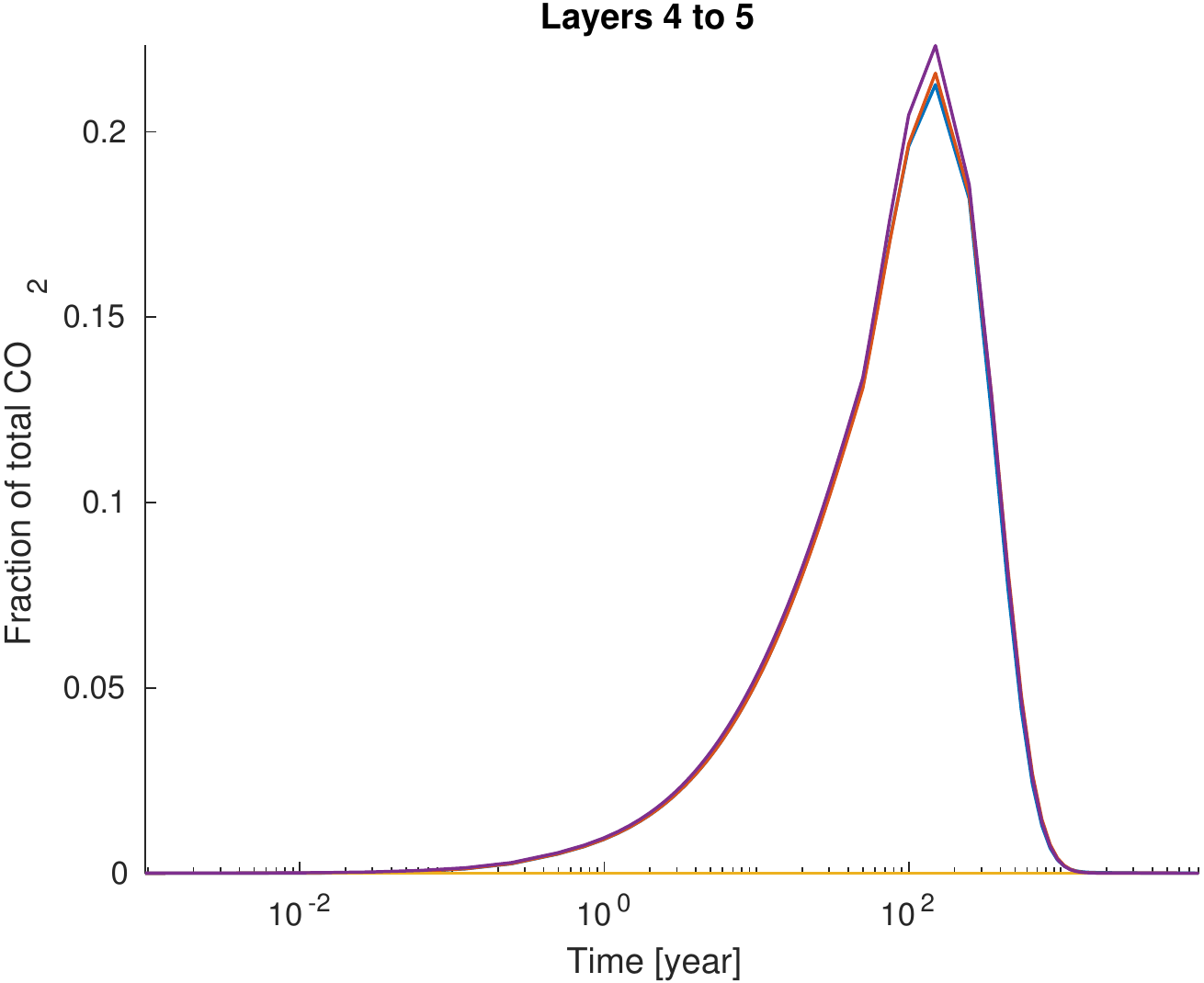}};
    \node[anchor=center,draw=black] at (2.5,-3.2) (s3) {%
      \includegraphics[width=7.0cm]{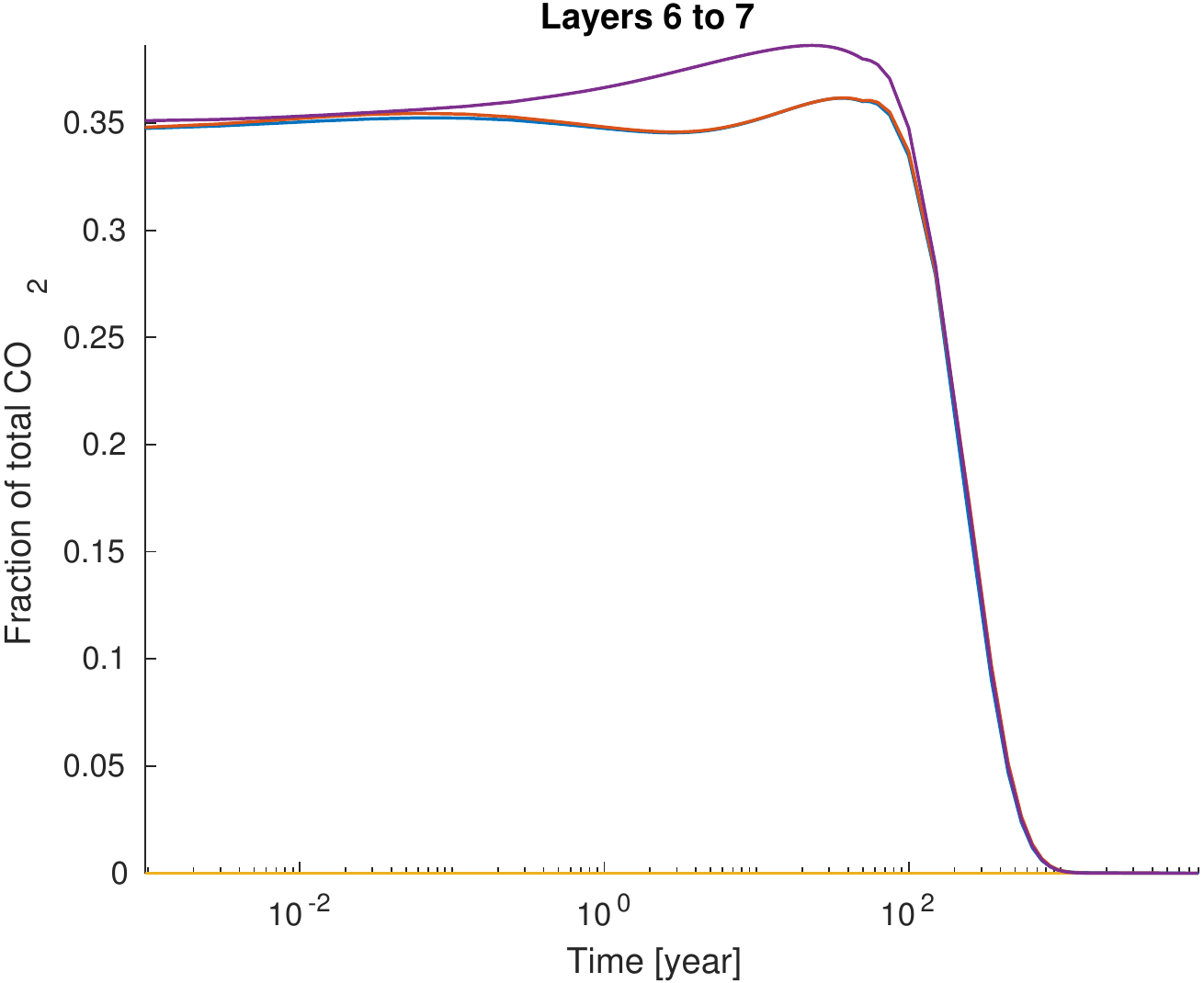}};
    \node[anchor=center,draw=black] at (10,-3.2) (s4) {%
      \includegraphics[width=7.0cm]{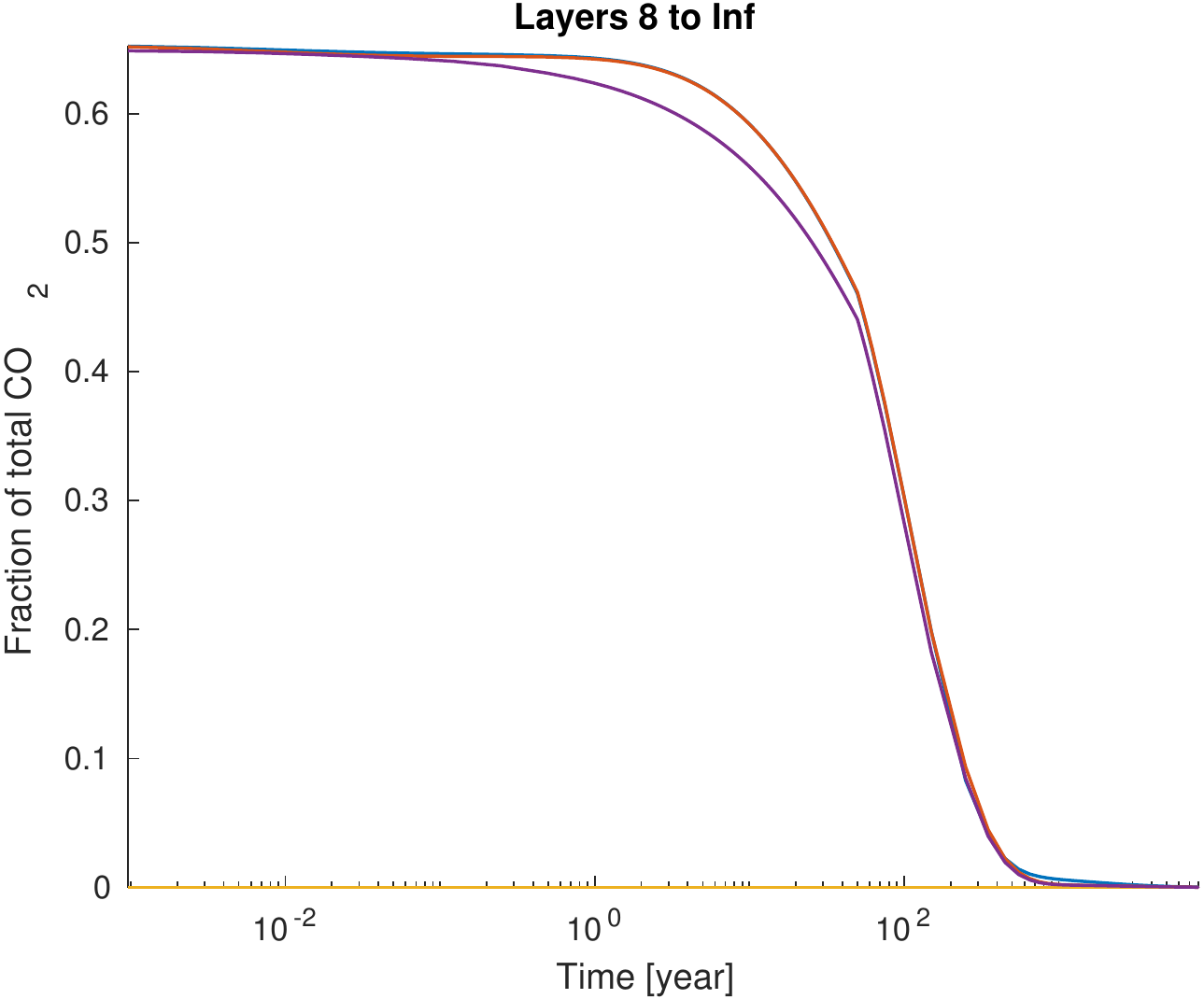}};
    \draw[->,black!30,  ultra thick,-latex] (s1.south) -- (3, 3.5);

    \draw[->,black!30, ultra thick,-latex] (s2.south) -- (4.9,1.8);

    \draw[->,black!30, ultra thick,-latex] (s3.north) -- (4.5,1.6);
    \draw[->,black!30, ultra thick,-latex] (s4.north) -- (4.5,0.8);
    \tiny
    \node[anchor=south east,red]   at (s1.north) {};
    \node[anchor=south east,black] at (s2.south east) {};
    \node[anchor=north east,blue]  at (s3.north east) {};
      \node[anchor=center] at (1.0,0.9) {%
        \includegraphics[width=2.8cm]{figs/sleipner_subset/legend_sleipner}};
  \end{tikzpicture}
  \caption{The Utsira model. Four different flow compartments are defined, with semi-permeable layers in-between shown in black. These shales cover the entire layer, but allow for diffusive leakage with a small rate compared to the vertical flow during injection.}
  \label{fig:utsira_migration}
\end{figure}

The final example is a 3D model with logical dimensions $64\times240\times15$ and 111,162 active cells. The geometry, permeability, well configuration and layering structure is taken from one of the Utsira models considered in \cite{utsira_pham}. The model has a uniform permeability of 1000 mD, with three semi-permeable layers as shown in Figure~\ref{fig:utsira_migration}. The corresponding faces in the vertical connections has a transmissibility multiplier of $10^{-4}$. 140.89 megatonnes of \co is injected over 50 years, followed by a migration period until 8000 years in total has passed. The boundaries of the reservoir is considered to be closed. Due to the coarseness of the vertical layers and fairly long time-steps, we consider linear relative permeability curves to be a reasonable choice to ensure correct segregation speed.

Unlike the previous example, the fine-scale model does not have sufficient resolution to be considered fully resolved. It is therefore not necessarily given that the fine-scale model is more accurate than the different VE solvers, which represent the plume depth continuously within each column. Four models are considered: The fine-scale model (111,162 cells), VE without layering (11,680 cells), VE with layers (42,877 cells) and hybrid-VE with fine-scale near the wells (44,215). The ratio between fine-scale and VE models is lower in this example, as the original model is already very coarse. In this sense, the VE models do not significantly reduce the number of cells, but they potentially increase the accuracy of the predicted \co distribution.

\begin{figure}[htbp]
	\centering
	\subfloat[End of injection, fine-scale]{
		\includegraphics[width=.4\textwidth]{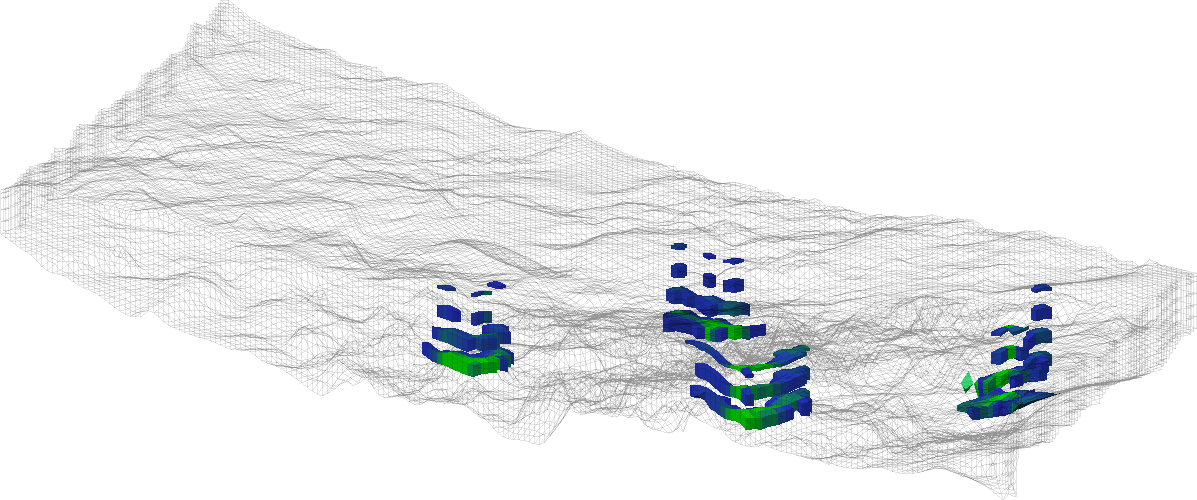}} \hspace{.25cm}
	\subfloat[End of migration, fine-scale]{
		\includegraphics[width=.4\textwidth]{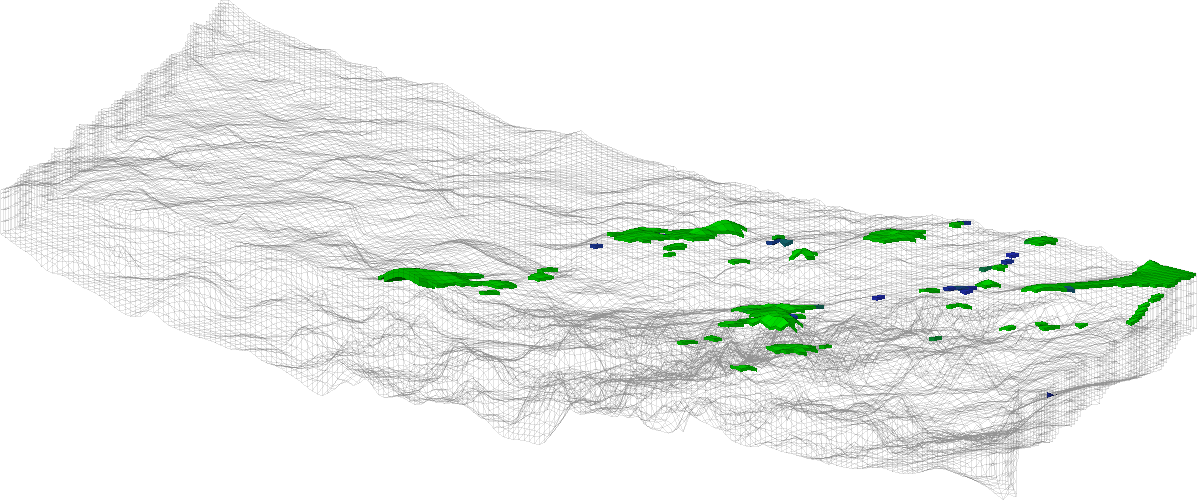}} \hspace{.25cm}\\
	\subfloat[End of injection, hybrid]{
		\includegraphics[width=.4\textwidth]{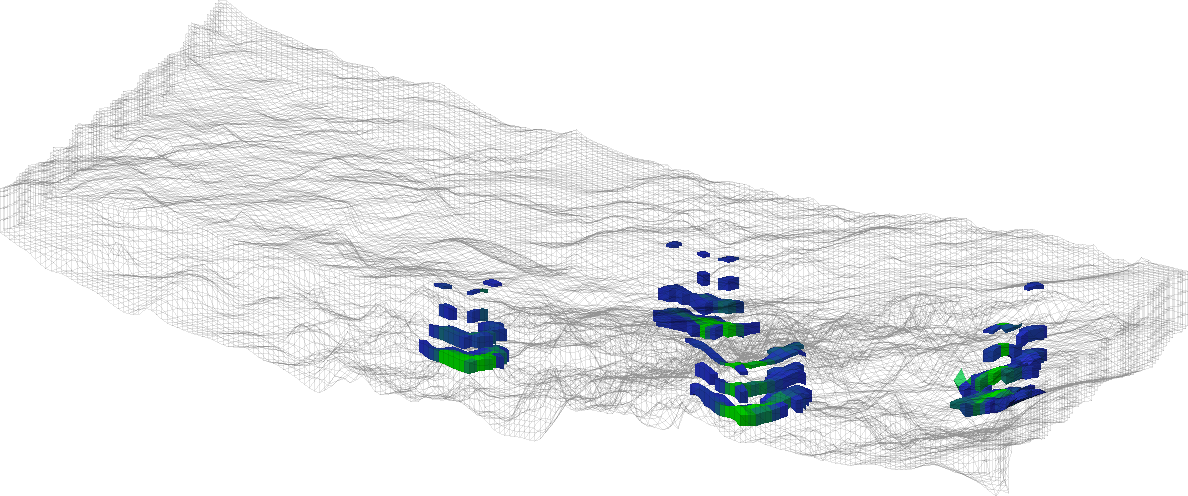}} \hspace{.25cm}
	\subfloat[End of migration, hybrid]{
		\includegraphics[width=.4\textwidth]{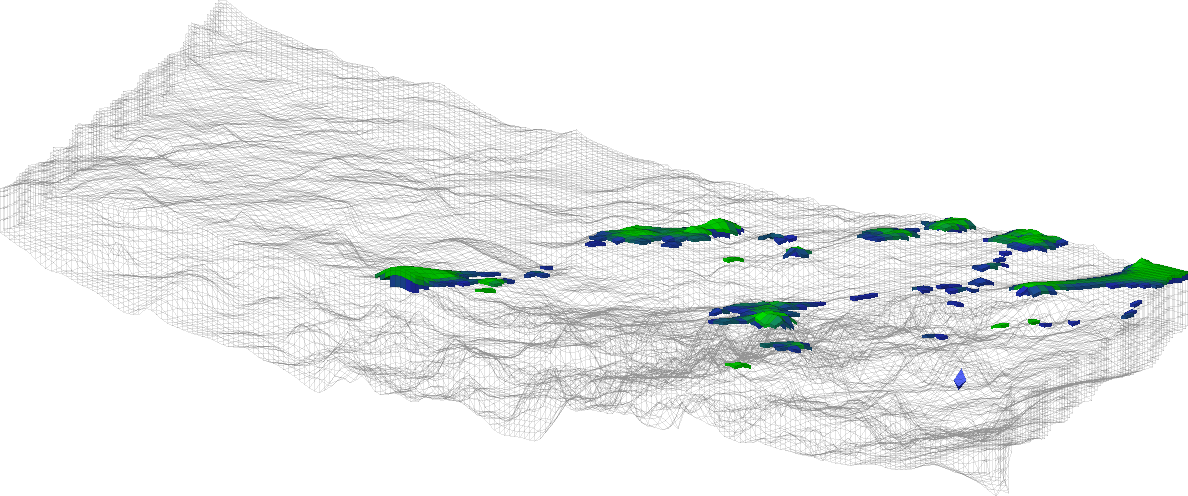}} \hspace{.25cm}\\
	\subfloat[End of injection, vertical-equilibrium]{
		\includegraphics[width=.4\textwidth]{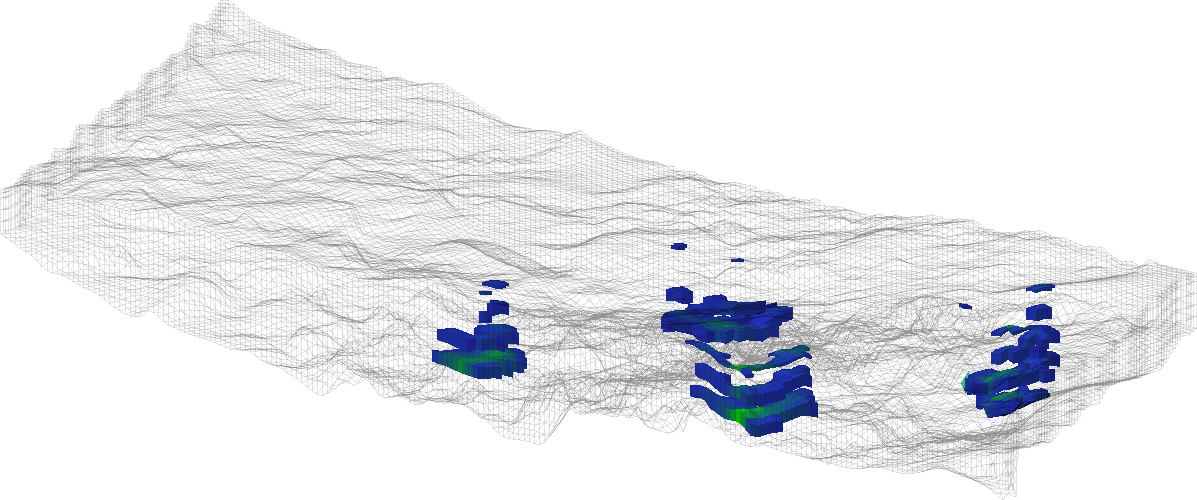}} \hspace{.25cm}
	\subfloat[End of migration, vertical-equilibrium]{
		\includegraphics[width=.4\textwidth]{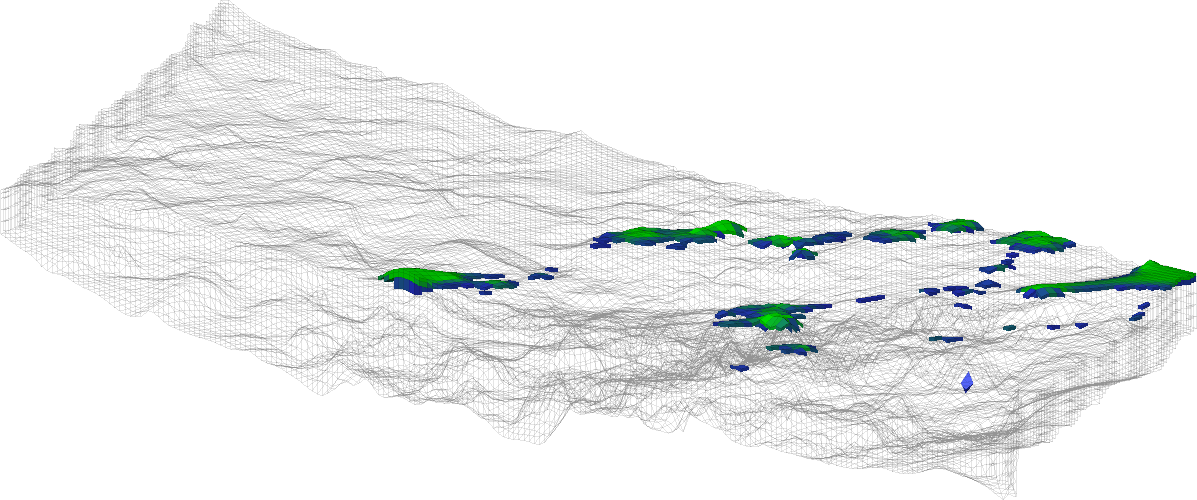}} \hspace{.25cm}\\
	\subfloat[End of injection, vertical-equilibrium (without layers)]{
		\includegraphics[width=.4\textwidth]{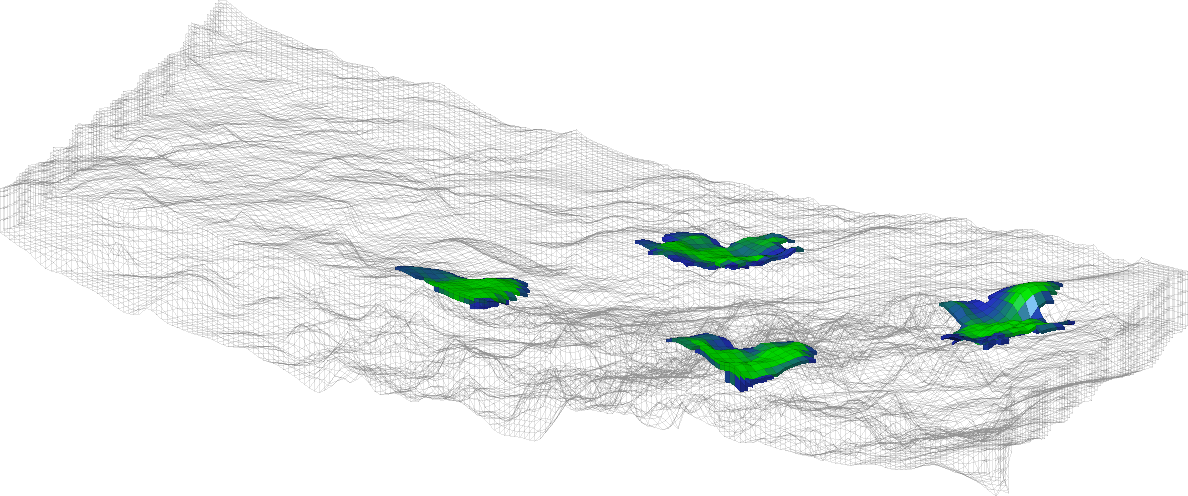}} \hspace{.25cm}
	\subfloat[End of migration, vertical-equilibrium (without layers)]{
		\includegraphics[width=.4\textwidth]{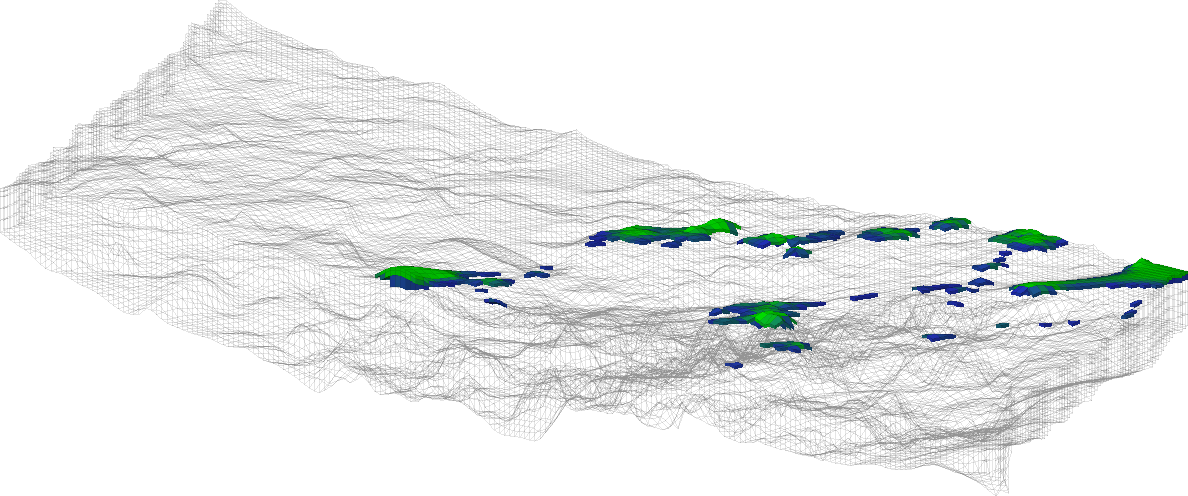}} \hspace{.25cm}\\

	\caption{Gas saturation in the Utsira example. A thresholding of $s_g > 0.1$ was used to visualize the plume.}
	\label{fig:utsira_sat}
\end{figure}

\begin{figure}[htbp]
	\centering
	\subfloat[BHP, injector 1]{
		\includegraphics[width=.5\textwidth]{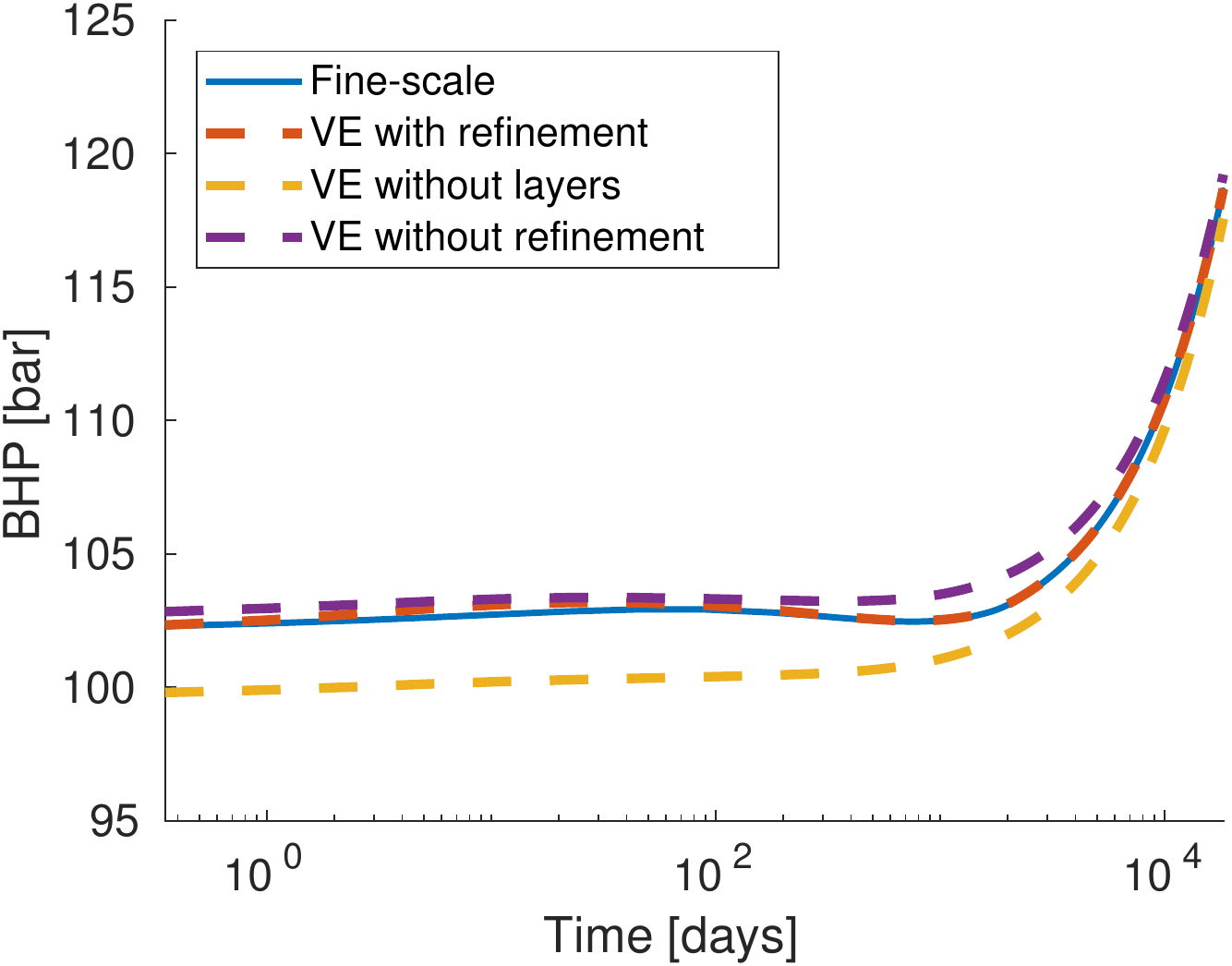}}
	\subfloat[BHP, injector 2]{
		\includegraphics[width=.5\textwidth]{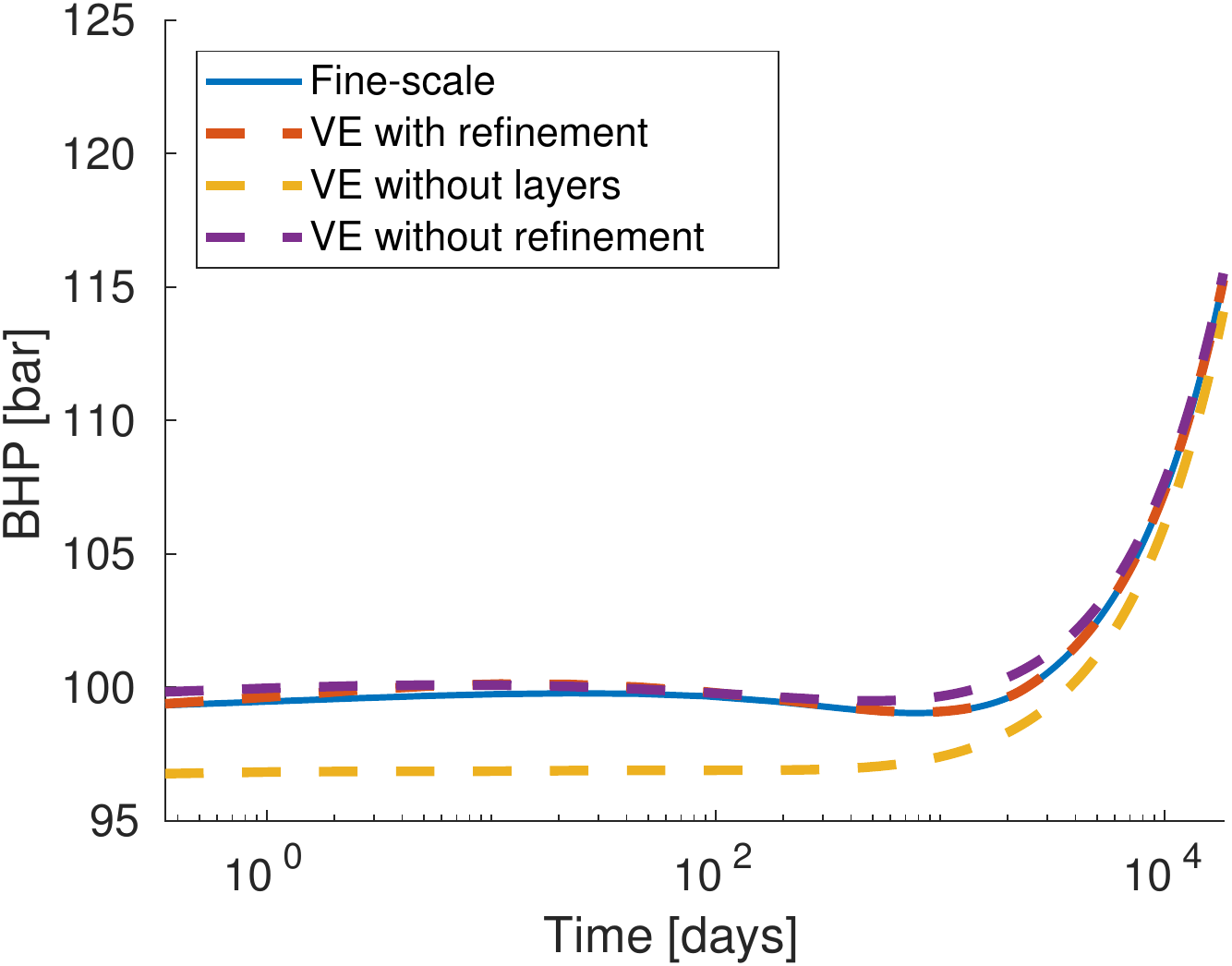}}\\
	\subfloat[BHP, injector 3]{
		\includegraphics[width=.5\textwidth]{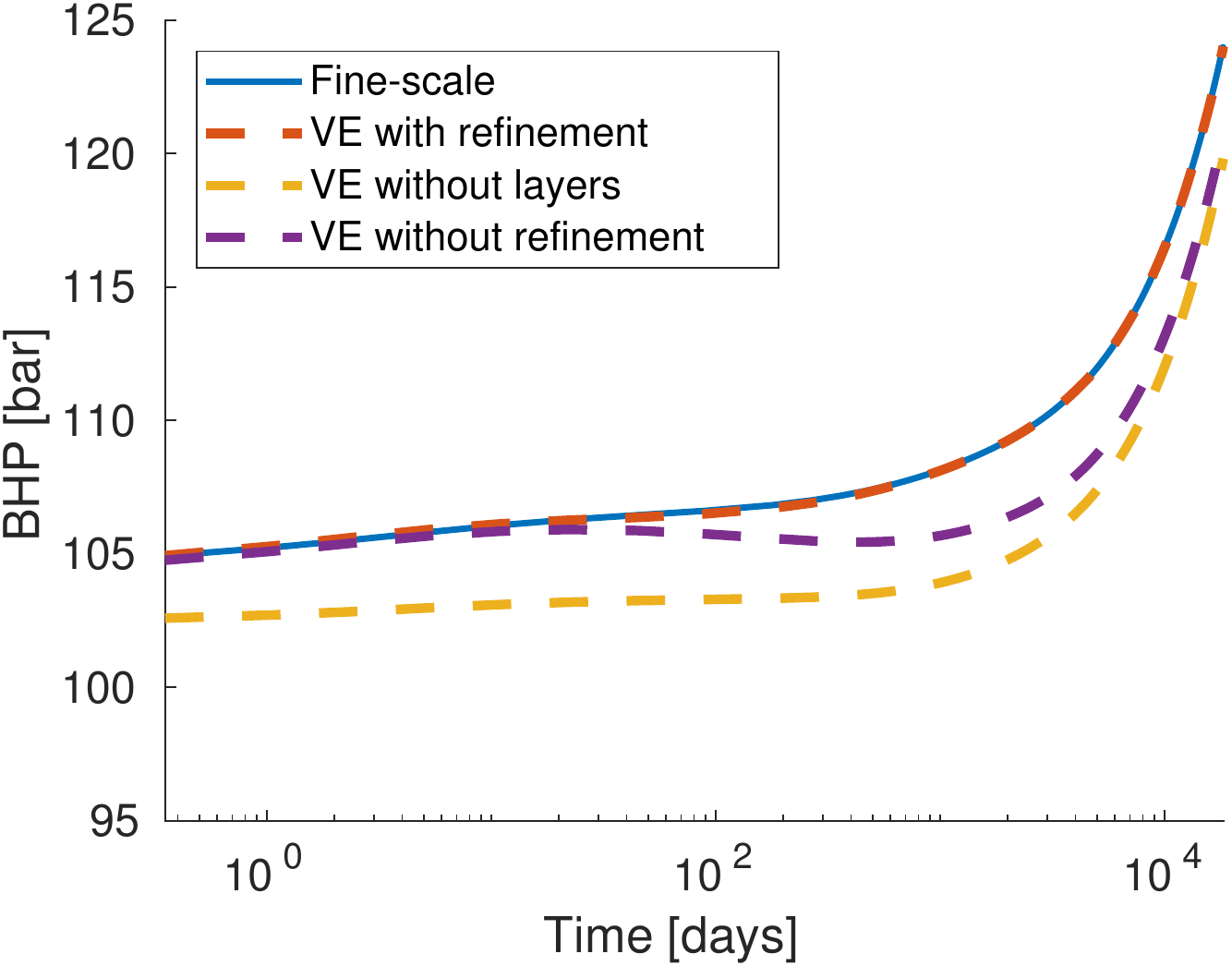}}
	\subfloat[BHP, injector 4]{
		\includegraphics[width=.5\textwidth]{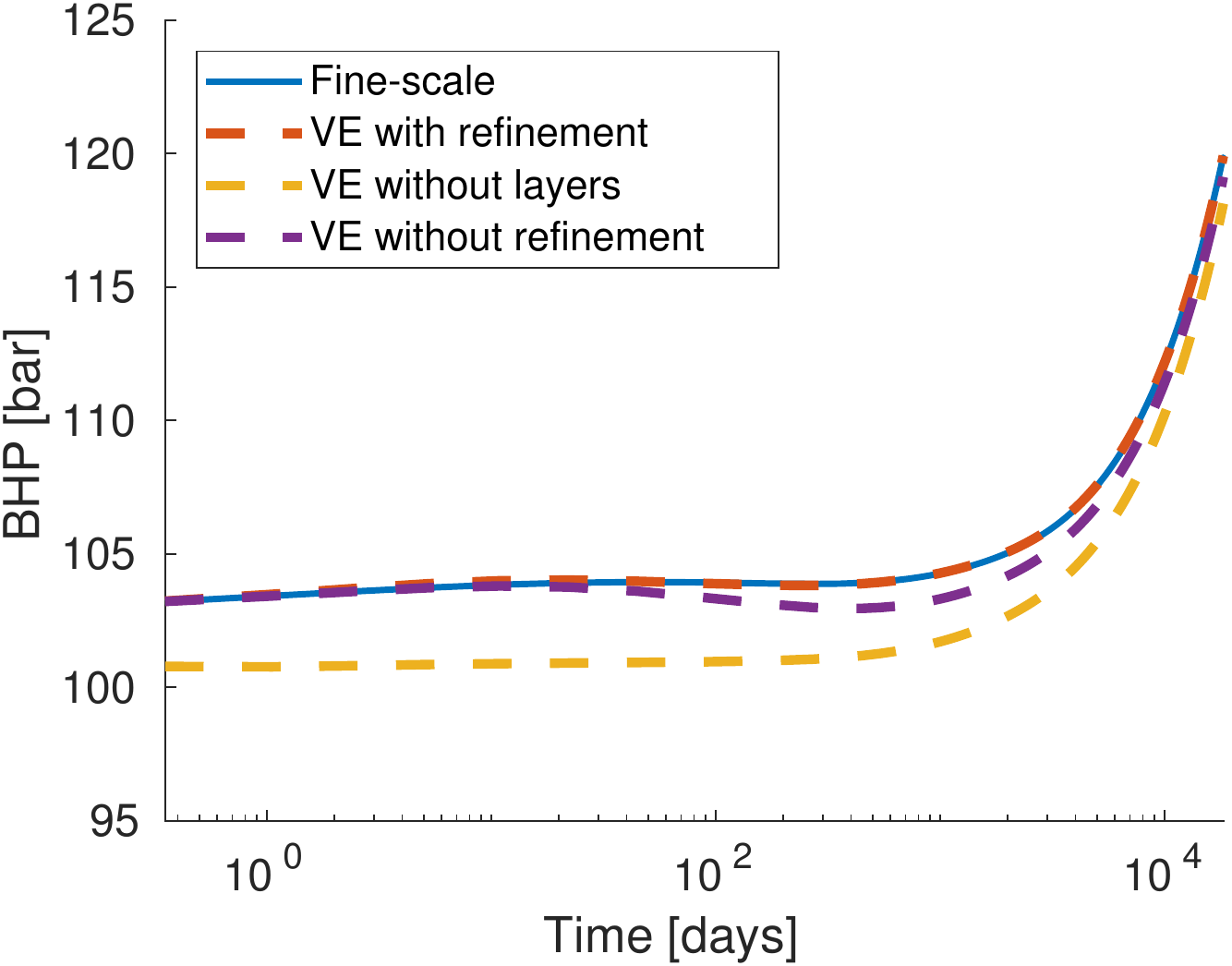}}
	\caption{Bottom hole pressures for the injectors in the Utsira example during the injection period}
	\label{fig:utsira_bhp}
\end{figure}

From the thresholded saturation profiles in Figure~\ref{fig:utsira_sat}, we observe that all four models give essentially the same trapping structure after 8000 years. As only the top surface is truly impermeable, the long term migration is less impacted by the inclusion of the layers. In the short and medium-term in Figure~\ref{fig:utsira_migration}), however, we note that the inclusion of layers is necessary to estimate which depths of the reservoir contains mobile \co. We also observe that the hybrid model exactly matches the fine-scale distribution, while the layered model without near-well refinement deviates somewhat. We emphasize that the hybrid model will represent the near-well flow exactly as in the fine-scale model, including any discretization error made by using large grid blocks. This is again reflected in the well curves, shown in Figure~\ref{fig:utsira_bhp}, where the hybrid model accurately predicts BHP.

\begin{figure}[htbp]
	\centering
	\subfloat[Sleipner subset]{
		\includegraphics[width=.5\textwidth]{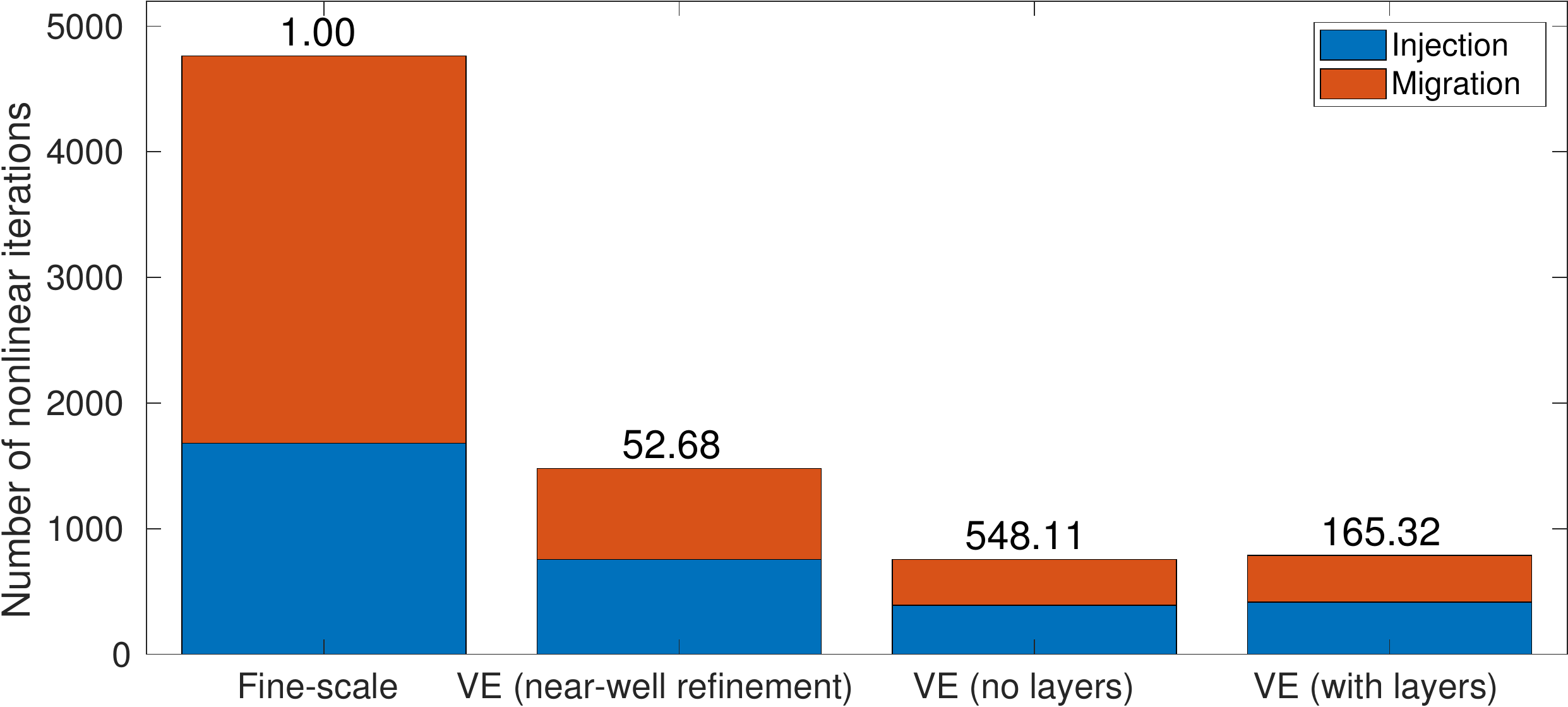}}
	\subfloat[Utsira model]{
		\includegraphics[width=.5\textwidth]{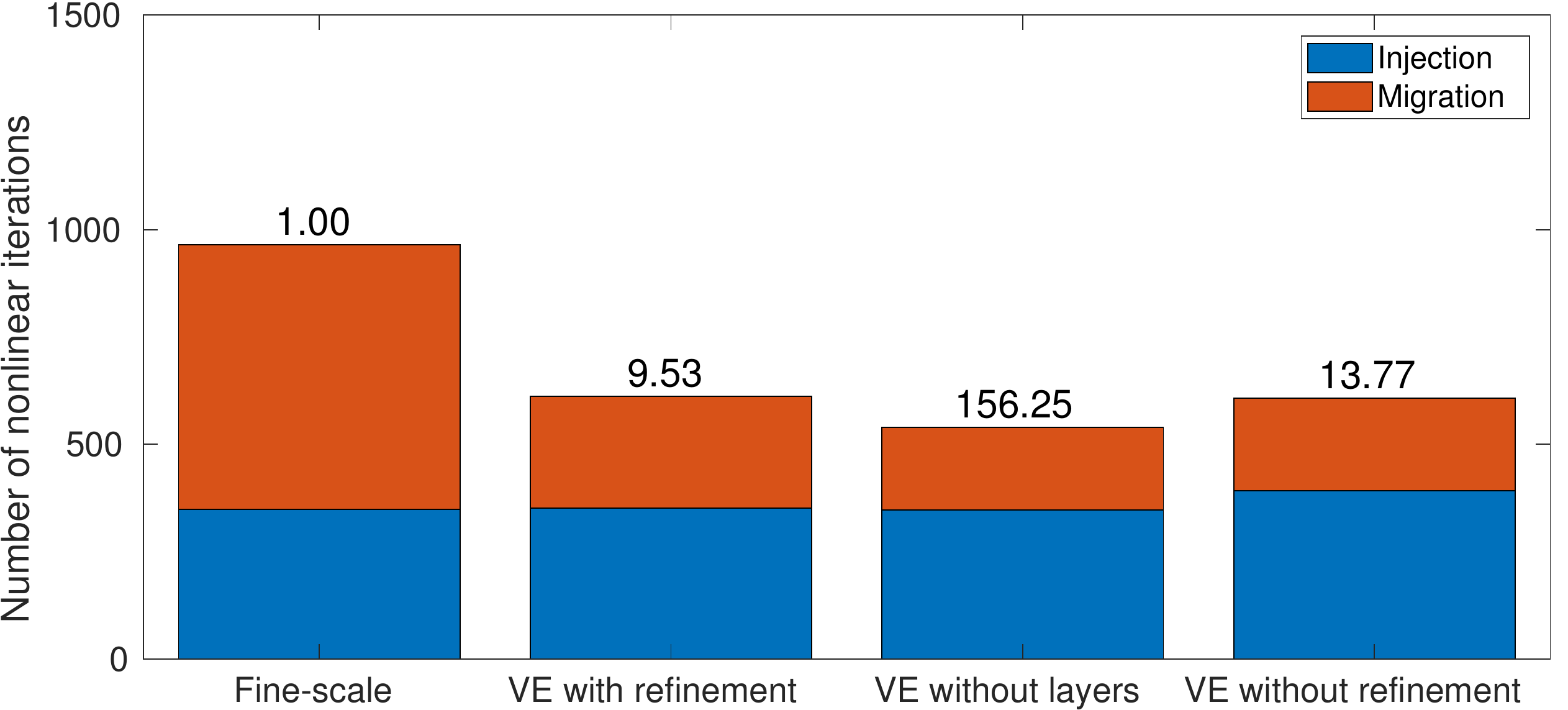}}\\
	\caption{Nonlinear iterations used in the Sleipner and Utsira examples. The number of iterations spent is divided between injection and migration, and relative speed-up compared to the fine-scale is shown as a number over each column.}
	\label{fig:iterations}
\end{figure}

The right subfigure in Figure~\ref{fig:iterations} contains the iteration numbers and relative speedups. As the model is already fairly coarse, the effect of applying VE is not as significant as in the Sleipner subset example. However, we still observe an order-of-magnitude speedup and a halving of the number of nonlinear iterations when compared to the fine-scale.

\FloatBarrier
\section{Conclusion}
We have presented a uniform framework for coupling VE models with general layered structure with traditional 3D reservoir models. In addition we have devices a general strategy for obtaining the model from a underlaying 3D reservoir model in industry standard format. In doing so we have introduced to our knowledge introduced new coupling terms between 3D models and VE model, between VE layers and a new diffuse leakage calculation consistent with the underlying 3D model. The model is implemented in the MRST AD-OO framework with state of the art robust implicit TPFA-MUP discretization including possibilities for adjoint calculations. This makes the model presented in this paper suited for use in optimization and data integration applications.

A major contribution of the paper is the uniform representation of the model in terms of pseudo relative permeabilities and capillary pressure, which highlights the need for relative permeabilities associated with oriented faces for upscaled model. This can bee seen as a discrete representation of general tensor mobilities.

 VE models which have shown great promise for evaluating long-term aquifer-scale \co storage and this work is an important step towards applying these models to general industry-grade models with layered structure, complex well models and non-VE flow regions. As the discretization is based on a underlying 3D detailed description it is always possible to check the validity of a coarse model by adding more parts of the model in the 3D region. The proposed treatment results in a practical and efficient simulation model for optimization and data integration in the setting of \co storage. However, we believe the methodology is well suited for other sub-surface simulation problems where segregated flow dominates large parts of the domain, for example gas storage and petroleum reservoirs with large gravity differences and large distances between wells.

 The model including examples of this paper will be made public under the GPLv3 with the next release of the Matlab Reservoir Simulation Toolbox \cite{MRST:2011a}.

\section{Acknowledgements}
This work was funded in part by the Research Council of Norway through grant no.\ 243729
(Simulation and optimization of large-scale, aquifer-wide \co injection in the North Sea) and
NCCS -- Industry Driven Innovation for Fast Track CCS Deployment grant no. 257579. Olav M{\o}yner is funded by VISTA, which is a basic research programme funded by Statoil and conducted in close collaboration with The Norwegian Academy of Science and Letters.

Statoil and the Sleipner License are acknowledged for provision of the Sleipner 2010
Reference dataset. Any conclusions in this paper concerning the Sleipner field are the
authors' own opinions and do not necessarily represent the views of Statoil.

We also acknowledge The Norwegian Petroleum Directorate for providing the model accompanying the paper \cite{utsira_pham} used in Example 4.

\clearpage

\bibliographystyle{plain}
\bibliography{bib/verticalaverage,bib/co2,bib/coarsen,bib/divSINTEF,bib/upscaling,bib/mrst,bib/polymer,bib/misc,bib/streamlines,added_refs,extra}
\end{document}